\documentclass{emulateapj}
\slugcomment{{\sc Accepted to ApJ} May 25, 2010}

  \def\etal{{\it et al.\ }}
\def\eg{{\it e.g.\ }}   
   
   \def\p3m{P${}^3$M}
\def\ap3m{AP${}^3$M} \def\-{{\em{---}}} 

\newcommand{\be}{\begin{equation}}  \newcommand{\ba}{\begin{eqnarray}}
\newcommand{\ee}{\end{equation}}  \newcommand{\ea}{\end{eqnarray}}

 \newcommand{\bi}{\begin{itemize}}
\newcommand{\ei}{\end{itemize}}  
  
 \newcommand{\Myr}{\,\textrm{Myr}}

\def\lesssim{\mathrel{\hbox{\rlap{\hbox{\lower4pt\hbox{$\sim$}}}\hbox{$<$}}}}
\def\gtrsim{\mathrel{\hbox{\rlap{\hbox{\lower4pt\hbox{$\sim$}}}\hbox{$>$}}}}

\usepackage{amsmath}

\begin{document}


\shorttitle{Gray \& Scannapieco} \shortauthors{}

\title{Formation of Compact Stellar Clusters by High-Redshift Galaxy Outflows I: Nonequillibrium Coolant Formation}

\author{William J Gray\altaffilmark{1}, \& Evan
  Scannapieco\altaffilmark{1}}

\altaffiltext{1} {School of Earth and Space Exploration,  Arizona
  State University, P.O.  Box 871404, Tempe, AZ, 85287-1404.}

\begin{abstract}

We use high-resolution three-dimensional adaptive mesh refinement
simulations to investigate the interaction of high-redshift
galaxy outflows with low-mass virialized clouds of primordial
composition. While atomic cooling allows star formation in objects
with virial temperatures above $10^4$ K, ``minihaloes" below this
 threshold are generally unable to form stars by
themselves. However, these objects are highly susceptible to triggered 
star formation, induced by outflows from neighboring high-redshift starburst
galaxies.
Here we conduct a study of these interactions, focusing on
cooling through non-equilibrium
molecular hydrogen (H$_2$) and hydrogen deuteride (HD)
formation. Tracking the non-equilibrium chemistry and cooling of 14
species and including the presence of a dissociating background,
we show that shock interactions can transform minihaloes into
extremely compact clusters of coeval stars.  Furthermore, these clusters are all
less than $\approx 10^6 M_\odot,$ and they are ejected from their parent dark matter halos:
properties that are remarkably similar to those of the old population of
globular clusters. 
	
\end{abstract}

 
\section{Introduction}

A generic prediction of the cold dark matter model is a large
high-redshift population of gravitationally-bound clouds  that are
unable to form stars.  Because atomic H and He line cooling is only effective
at temperatures above $10^4$ K, clouds of gas and dark
matter  with virial temperatures below this threshold must radiate
energy through dust and molecular line emission.   While the levels of
H$_2$ left over from recombination are sufficient to cool gas in the
earliest structures (\eg Abel \etal 2002; Bromm \etal 2002), the
resulting 11.20 - 13.6 eV background emission from the stars in these
objects (\eg Haiman \etal 1997; Ciardi \etal 2000; Sokasian \etal
2004; O'Shea \& Norman 2007) is likely to have quickly dissociated
these trace levels of primordial molecules (Galli \& Palla 1998).   And although an early
X-ray background could have provided enough free electrons to promote
H$_2$ formation, the relative strength between these two backgrounds is
uncertain, and it is unlikely that the background was strong enough to
balance ultraviolet (UV) photodissociation. Even if there were some trace amount of
H$_2$ in these clouds, it is likely to be in such a small abundance as to not
impact their structure (Whalen \etal 2008a; Ahn \etal 2009). 

The result is a large number of dark matter halos that were massive
enough to overcome the thermal pressure of the primordial intergalactic
medium and retain their gas,  but not massive enough to excite
the radiative transitions necessary to cool this gas into stars.  These
``minihalos,'' whose virial temperatures were $T_{\rm{vir}} \lesssim 10^{4}$ K 
and whose total masses were between $10^{4}$ and $10^{7.5} M_{\odot}$
would have remained sterile until some outside force either disrupted
them, or more interestingly, disturbed them so as to catalyze coolant
formation.  In fact, two possible coolant formation methods have
been considered in detail in the literature: ionization
fronts and shock fronts. 

In the case of ionization fronts, such as would occur during the
epoch of reionization, high-energy photons emitted from galaxies
or quasars interact with the neutral atomic
minihalo gas.  Bond, Szalay \& Silk (1988) originally  discussed how
the resulting photoionization  would expel the gas contained in a minihalo by
suddenly heating it to $T \approx  10^4$ K, as would be the case
in the optically-thin limit.  On the other hand,  Cen (2001), used
simple analytic estimates to argue that ionization fronts would cause
non-equilibrium H$_2$ formation and the collapse of the gas inside the
gravitational potential.  However, Barkana \& Loeb (1999) studied
minihalo evaporation using static models of uniformly illuminated
spherical clouds, accounting for optical depth and self-shielding
effects, and showed that the cosmic UV background boiled most
of the gas out of these objects.  Later, Haiman, Abel, \& Madau (2001) carried out
three-dimensional (3D) hydrodynamic simulations  assuming the minihalo  gas
was spontaneously heated to $10^4$ K,  also finding quick disruption.
Finally, full radiation-hydrodynamical simulations of ionization
front-minihalo interactions were carried out in Iliev \etal (2005) and
Shapiro \etal (2004;  see also Shapiro, Raga \& Mellema 1997; 1998).
These demonstrated that intergalactic ionization fronts 
decelerated when they encountered the dense, neutral
gas inside minihaloes and were thereby transformed into D-type
fronts, preceded by shocks that completely photoevaporated the
minihalo gas. 

A second and more promising avenue for coolant formation is the
interaction between galactic outflows and minihaloes. These
galaxy-scale winds, which are driven by core-collapse supernova
and winds from massive stars, are commonly observed around dwarf and
massive starbursting galaxies at both low and high redshifts  (\eg
Lehnert \& Heckman 1996; Franx \etal 1997; Pettini \etal 1998;
Martin 1999; 1998; Heckman \etal 2000; Veilleux \etal 2005; Rupke
\etal 2005), and a variety of theoretical arguments suggest
that these galaxies represent only the tail end of a larger population
of smaller  ``pre-galactic," starbursts that formed before reionization
(Scannapieco, Ferrara \& Madau 2002; Thacker, Scannapieco,
\& Davis 2002). Furthermore, the interstellar gas swept up in a starburst-driven
wind can effectively  trap the ionizing photons behind it (Fujita
\etal 2003), meaning that at high redshifts, many intergalactic
regions may have been impacted by outflows well before they were ionized.

Such $\approx 100-300$ km/s shocks can cause intense cooling
through two mechanisms: (i) the mixing of metals with ionization potentials below 13.6
eV (Dalgarno \& McCray 1972), which allow for atomic line cooling even at temperatures below $10^4$ K;
 and (ii) the formation of H$_2$ and HD by nonequilibrium processes (Mac Low \& Shull 1986; Shapiro \& Kang
1987; Kang \etal 1990; Ferrara 1998; Uehara \& Inutsuka 2000), which allow for molecular line cooling associated vibrational
 and rotational transitions (Palla \& Zinnecker
1988).  In fact, Scannapieco \etal (2004) showed that these effects
were so large that shock interactions could induce intense bursts of
cooling and collapse in previously ``sterile'' minihalo gas. Using
 simple analytic models, they found that the most likely result was
the formation of compact clusters of coeval stars, although they also
emphasized the importance of multidimensional numerical studies to
confirm this result.  

In this paper, we present the first in a series of 3D numerical
studies to capture  the physical interactions between 
primordial minihalos and supernova-driven galactic outflows.  While
triggered star formation has been simulated in low-redshift intergalactic clouds impacted
by radio jets (\eg van Breugel \etal 1985; Fragile \etal 2004; Wiita
2004; Klamer \etal 20004), this has never been numerically simulated
in the high-redshift, minihalo case.   In this paper, 
we  focus on the physics of non-equilibrium H$_2$ and HD chemistry and associated
cooling in determining the properties of the post-shock minihalo
gas.  In particular, we show that molecule formation and the resulting
cooling is strong enough to induce rapid minihalo collapse and star formation,
leading to compact stellar clusters.  

The structure of this paper is as follows.  In \S 2 we outline the
physical components of the model, focusing on our primordial H$_2$ and
HD chemical network and cooling routines and their respective
tests.  In \S 3 we outline the general model used for the galactic
outflow and the minihalo, and in \S 4 we present the simulation results and
relate the resulting stellar clusters to local observations.
Conclusions are given in \S 5.

\section{Numerical Method}

All simulations were performed with  FLASH version 3.1,
a multidimensional adaptive mesh refinement hydrodynamics
code  (Fryxell \etal 2000) that solves the Riemann problem on a
Cartesian grid using a directionally-split  Piecewise-Parabolic Method
(PPM) solver (Colella \& Woodward 1984; Colella \& Glaz 1985; Fryxell,
M\" uller, \& Arnett 1989).  Furthermore, unlike earlier versions of the
code, FLASH3 includes an effective parallel multigrid gravity solver as
described  in Ricker (2008). However, before shock-minihalo
interactions could be simulated, two capabilities needed to
be added: nonequilibirum primordial chemistry, and cooling from atoms 
and from molecules produced in these interactions.   In this section we describe
our numerical implementation of each of these processes, along with
the  tests we carried out before  applying the code to 
shock-minihalo interactions.

\subsection{Chemistry}

As the minihalos we are considering in this paper
are made up of primordial gas, their chemical makeup is highly restricted, with contributions from only
hydrogen, helium, and low levels of deuterium. Yet even these three 
isotopes can exist in a variety of ionization states and molecules and
are thus associated with a substantial network of chemical reactions that must be tracked 
throughout our simulations.
	
\subsubsection{Implementation}

The chemical network that was implemented into FLASH is outlined by
Glover \& Abel (2008, hereafter GA08).
Throughout our simulations we track three states of atomic hydrogen
(H, H$^+$, \& H$^-$) and  atomic deuterium (D, D$^+$, \& D$^-$), 
three states of atomic helium (He, He$^+$, $\&$, He$^{++}$),
two states of molecular hydrogen (H$_2$ $\&$ H$_2^+$) and molecular hydrogen
deuteride (HD $\&$ HD$^+$), and electrons (e$^-$). 
For simplification, any reaction that involved molecular
deuterium (D$_2$) and all three-body reactions were neglected. As
stated in GA08, the very small amount of D$_2$ and D$_2^+$ produced makes
any cooling by these molecules irrelevant, while three-body reactions
only become important at $n \gtrsim 10^8$ cm$^{-3}$  (\eg Palla \etal 1983), 
many orders of magnitude denser than the conditions considered here. With these constraints, 
a total of 84 reactions were used out of the 115 described in GA08.

Photodissociation rates due to an external radiation field were also included as given in Glover \& Savin (2009). 
These rates are calculated assuming a $T_{\rm eff} = 10^5$ K blackbody source and their strength is quantified 
by the flux at the Lyman limit, $J(\nu_{\alpha}) = 10^{-21} J_{21}$ erg $\rm s^{-1}$ $\rm cm^{-2}$ 
$\rm Hz^{-1}$ $\rm sr^{-1}$.  Note that once H$_2$ and HD  are produced in sufficient quantities, some molecules 
are self-shielded from the background radiation. However,
for simplicity, we considered only the  case where there was no self-shielding, and thus our results place an 
upper limit on the effect of a dissociating background.
This process adds an additional 7 reactions for a total of 91 reactions 
in the chemical network.

In reactions that involve free electrons recombining with ions, there
are two possible choices for the reaction rate, depending on the
overall optical depth of the  cloud to ionizing radiation. In the
optically-thin case (Case A; Osterbrock 1989) ionizing photons emitted
during recombination are lost to the system, while in  the
optically-thick case (Case B), ionizing photons are  reabsorbed by 
neighboring neutral atoms, which have the effect of lowering the
recombination rates by  essentially not allowing recombination to the
ground state.    There are three reactions (${\rm{H}}^+ + {\rm e}^- \rightarrow {\rm H} +
\gamma$, $ {\rm He}^+ + {\rm e}^- \rightarrow {\rm He} + \gamma$,  and ${\rm He}^{++} + {\rm e}^- \rightarrow
{\rm He}^+ + \gamma$)  where this is a concern, and as we shall see below, in
all cases our clouds are optically thick, such that Case B
recombination rates are appropriate.

The binding energy from each species is also important in the total energy budget and on the evolution of the gas. In FLASH these are defined such that all of the neutral species (H, D, e$^{-}$, \& H) have binding energies equal to zero. As the gas is heated and the atomic species begin to ionize, the endothermic reactions remove the binding energy between the nucleus and electron(s) from the internal energy of the gas. For H and D this requires 13.6 eV, while the ionized states for Helium (He$^{+}$ and He$^{++}$) have ionization potentals of 24.5 eV and 79.0 eV respectively. H$^-$ and D$^-$ are only weakly bound and have similar binding energies of 0.75 eV. Finally H$_2$ and HD have binding energies of 4.4 eV,  and H$_2^{+}$ and HD$^{+}$ have binding energies of 10.9 eV, somewhat lower than the atomic species.

To describe the evolution of our 14 species, we enumerate them with an index
$i$ such that each has $Z_i$ protons and $A_i$ nucleons,
following the structure and syntax from Timmes (1999).
Next we consider a gas with a total mass density $\rho$  and temperature $T$ and  
denote the number and mass densities of the $i$th isotope as 
$n_i$ and $\rho_i,$ respectively. For each species we also define a mass fraction 
\begin{equation}
 X_i \equiv \rho_i/ \rho = n_i A_i/(\rho N_A),
 \end{equation}
 where $N_A$ is Avogadro's number, and we define the molar abundance of the $i$th species as
 \begin{equation}
 Y_i \equiv X_i/A_i = n_i/(\rho N_A),
 \end{equation}
 where conservation of mass is given by $\sum^N_i X_i = 1.$
 Each of the 14 species can then be cast  as  a continuity equation in the form 
 \begin{equation}
\dot Y_i \equiv \frac{dY_i}{dt} = \dot{R_i},
 \end{equation}
where $\dot{R_i}$ is the total reaction rate due to all the binary reactions of the form $i+j \rightarrow k+l,$ defined as
\begin{equation}
\dot{R_i} \equiv \sum_{j,k} Y_l Y_k \lambda_{kj}(l) - Y_i Y_j \lambda_{jk}(i),
\label{rate_eqn}
\end{equation}
where $\lambda_{kj}$ and $\lambda_{jk}$ are the creation and destruction chemical reaction rates for a given species.
If the species in question is affected by UV background radiation, the continuity equation takes the following form,
\begin{equation}
\dot{R_i} \equiv \sum_{j,k} Y_l Y_k \lambda_{kj}(l) - Y_i Y_j \lambda_{jk}(i) - Y_i J(\nu_{\alpha}),
\label{rate_eqn2}
\end{equation}
where the last term accounts for the amount of these species that are destroyed by the background radiation, J$(\nu_{\alpha})$.
Throughout our simulations, changes in the number of free elections
are not calculated directly, but rather at the end of each cycle
we use charge conservation to calculate their molar fraction, as
\begin{equation}
Y_{elec} = Y_{{\rm H}^+} + Y_{{\rm D}^+} + Y_{{\rm HD}^+} + Y_{{\rm H}_2^+} + Y_{{\rm He}^+} + 2 Y_{{\rm He}^{++}} - Y_{{\rm H}^-} - Y_{{\rm D}^-}. 
\end{equation}

Because of the often complex ways that the chemical reaction rates depend on temperature and the intrinsic order of magnitude spread in the rates, the resulting equations are `stiff,'  meaning  that the ratio of the minimum and maximum eigenvalue of the Jacobian matrix, $ J_{i,j} = \partial \dot Y_i / \partial Y_j$, is large and imaginary. This means that implicit or semi-implicit methods are necessary to efficiently follow their evolution.
To address this problem, we arrange the molar fractions of the 13 species, excluding e$^-$, into a vector ${\bf Y,}$
and solve the resulting system of equations using a $4^{th}$ order accurate Kaps-Rentrop, or Rosenbrock method
(Kaps \& Rentrop 1979). In this method, the network is advanced over a time step $h$ via
\begin{equation}
{\bf Y}^{n+1} = {\bf Y}^{n} + \sum^{4}_{i=1} b_i {\bf \Delta}_i, 
\end{equation}
where the  $\bf \Delta_i$ vectors are found by successively solving the four matrix equations
\begin{align}
(\hat{1}/\gamma h - \bar{J}) \cdot {\bf \Delta}_1 &= f({\bf Y}^n), \\
(\hat{1}/\gamma h - \bar{J}) \cdot {\bf \Delta}_2 &= f({\bf Y}^n+a_{21}{\bf \Delta}_1) + c_{21}{\bf \Delta}_1/h, \\
(\hat{1}/\gamma h - \bar{J}) \cdot {\bf \Delta}_3 &= f({\bf Y}^n+a_{31}{\bf \Delta}_1 + a_{32}{\bf \Delta}_2) \nonumber \\ & + (c_{31}{\bf \Delta}_1+c_{32}{\bf \Delta}_2)/h, \qquad {\rm and} \\
(\hat{1}/\gamma h - \bar{J}) \cdot {\bf \Delta}_4 &= f({\bf Y}^n+a_{41}{\bf \Delta}_1 + a_{42}{\bf \Delta}_2+a_{43}{\bf \Delta}_3) \nonumber \\ & + (c_{41}{\bf \Delta}_1+c_{42}{\bf \Delta}_2+c_{43}{\bf \Delta}_3)/h.
\end{align}

Here $b_i, \gamma, a_{ij}$, and $c_{ij}$ are fixed constants of the method, $f({\bf Y}) \equiv \dot {\bf Y},$ $\hat{1}$ is the identity matrix, and $\bar J$ is the Jacobian matrix.
Note that the four matrix equations represent a staged set of linear equations and that the four right hand sides are not known in advance. At each step, an error estimate is given for the difference between the third and fourth order solutions. 
For comparison we also carried out tests, using a multi-order Bader-Deuflhard method (Bader \& Deuflhard 1983). However in the end, 
the Rosenbrock method was chosen over this method because of its efficiency and speed.

As the species evolve, the temperature of the gas changes from the release of internal energy from recombinations or the loss of internal energy from ionizations and dissociations. These changes can 
in turn affect the reaction rates. Thus to ensure the stability of the chemistry routine 
while at the same time allowing the simulation to proceed at the hydrodynamic time-step,
 we developed a method of cycling over multiple Kaps-Rentrop time steps within a single hydrodynamic time step.
Here we estimated an initial chemical time step of each species as 
\begin{equation}
	\tau_{{\rm chem},i} = \alpha_{\rm chem} \frac{{Y}_i+0.1Y_{H^+}}{\dot{Y_i}}, 
\label{tau_chem}
\end{equation}
where $\alpha_{\rm chem}$ is a constant determined at runtime that controls the desired fractional change of the fastest evolving species. The change in molar abundances, $\dot{Y}_i$'s, were calculated from the ordinary differential equations that make up the chemical network, and the molar fractions of each species $Y_i$'s are given by the current values.  In both the tests and simulations, we chose a value of $\alpha_{\rm chem} = 0.5.$ 

Note that we offset the subcycling time step by adding a small fraction of the ionized hydrogen abundance to eq.\ (\ref{tau_chem}).  This is because
there are conditions where a species is very low in abundance but changing very quickly, for example, rapid ionization of  atomic species, which will cause the subcycling to run away with extremely small time steps.  In regions in which most species are neutral, this has little effect since the chemical time step is likely longer than the hydrodynamical one, and in regions in which abundances are rapidly changing, then this extra term buffers against very small times steps. It also prevents rapid changes in internal energy as energy is removed as atomic species are ionized and gained as they recombine. 
	
\begin{figure}[ht!]
\centering
\includegraphics[trim = 0mm 0mm 0mm 10mm,clip,scale=0.40]{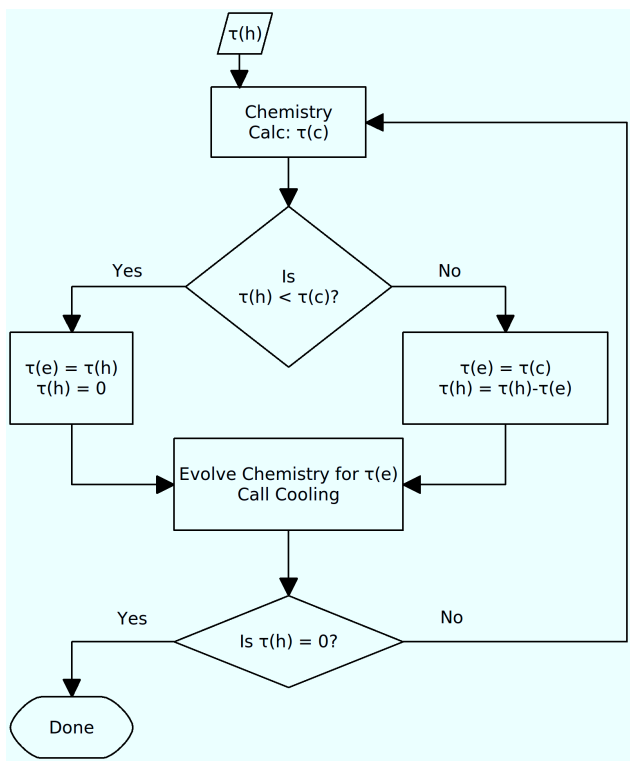}
\caption{Schematic view of chemistry subcycling. First, the chemical time step, $\tau(c),$ is calculated using eqn.~(\ref{tau_chem}). If this is larger than the hydrodynamic time step, then the evolution time step, $\tau(e),$ is set to the hydrodynamic time step. Else, $\tau(e)$ is set to the chemical time step. The network is then evolved for $\tau(e)$ and the remaining time step $\tau(h)$ is calculated. If this is zero then we proceed to the next step, else we cycle back through the network with the updated abundances. This loop continues until the full hydrodynamic time step is covered. Note that after every chemical network iteration, the cooling routine is called. }
\label{chem_flow}
\end{figure}

Once calculated, these species time steps are compared to each other and the smallest time step, associated with the fastest evolving species, is chosen as the subcycle time step. If this is longer than the hydrodynamic time step, the hydrodynamic time step is used instead and no additional subcycling is done. If subcycling is required, the species time step is subtracted from the total hydrodynamic time step and the network is then updated over the chemical time step. The species time steps are recalculated after each subcycle and compared to the remaining hydrodynamic time step. This is repeated until a full hydrodynamic time step is completed, as is schematically shown in Fig.~\ref{chem_flow}.

In cases in which the gas is extremely hot or cold, the chemical make-up can be determined directly from the temperature, avoiding the need for matrix inversions. If the temperature is above $10^{5}$ K then all atomic species become ionized and all molecular species are dissociated, and the network can be bypassed. If the temperature is between $2.0 \times 10^4$ and $1.0 \times 10^5$ K , then we `prime' the solutions and ionize 5$\%$ of available neutral hydrogen, 5$\%$ of neutral helium ($4.5\%$ into singly ionized helium and $0.5\%$ into doubly ionized helium), before entering the iterative solver, to help accelerate the routine towards the correct solution.  Finally if the temperature is less than 50 K, then all species are kept the same, and no reactions are calculated. This is done because cooling and chemistry rates become unimportant at such low temperatures. It also has the benefit of speeding up the simulation slightly as very little time is spent in either the cooling or chemistry routines. In all other cases, the full network is evolved without alterations.

\subsubsection{Chemistry Tests}

\begin{figure}[ht]
\centering
\includegraphics[trim = 0mm 0mm 0mm 0mm,clip,scale=0.45]{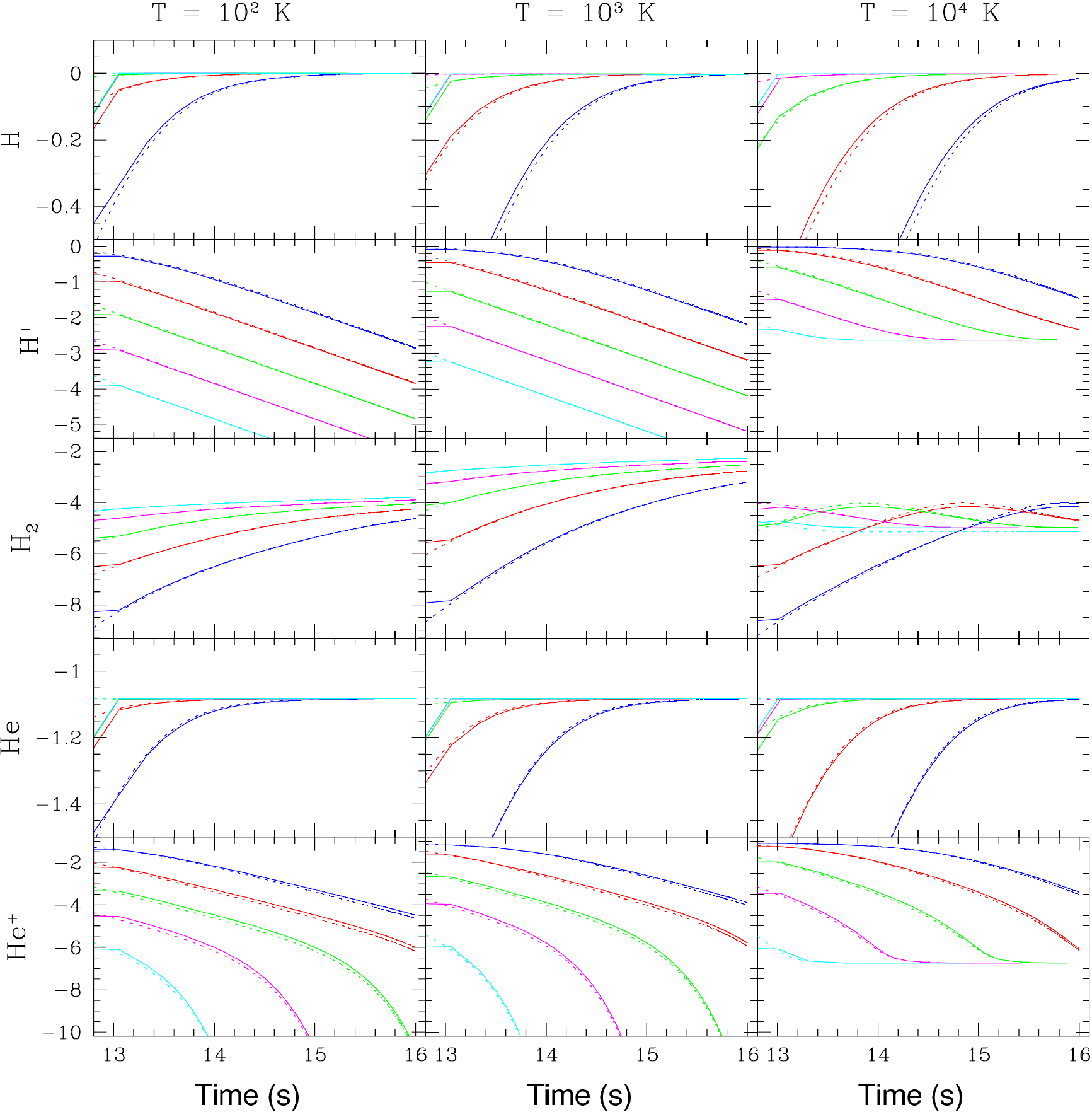}
\caption{Chemical evolution tests. Column 1 shows the $T=10^2 \rm \ K$ case, column 2 shows the $T=10^3 \rm \ K$ case, and column 3 shows the $T=10^4 \rm \ K$. Time is given on the $x$-axis and the number density of each species divided by the total number density of hydrogen is given on the $y$-axis. The blue lines correspond to the $n=0.01 \ {\rm cm}^{-3}$ case, red to the $n=0.1 \ {\rm cm}^{-3}$ case, green to the $n=1.0 \ {\rm cm}^{-3}$ case, magenta to the $n=10.0 \ {\rm cm}^{-3}$ case, and teal to the $n=100.0 \ {\rm cm}^{-3}$ case. The solid lines are results from FLASH and the dashed lines are results from G09.}
\label{ctest}
\end{figure}

To test our implemented chemical network, we carried out a series of runs in which initially dissociated and ionized gas was held at constant temperature and density for $10^{16}$ seconds.  A small initial time step ($t_0 \approx$ $10^6$ s) was used and allowed to increase up a maximum time step of $10^{12}$ seconds. 
Models were run with total hydrogen number densities varying from $0.01 \ \rm cm^{-3}$ to $ 100 \ \rm cm^{-3}$, and temperatures ranging between $10^2 \rm \ K$ and $10^{4} \rm \ K$, and no external radiation.
In each case, the results were compared to the results of a different implementation of the same chemical network within the ENZO code  (Glover 2009, private communication, G09), yielding the
molar fractions
 shown in Figure~\ref{ctest}. In this figure, the three columns correspond to runs with  different temperatures, the curves corresponds to runs with different densities, and
the rows correspond to the evolution of different species.  

The match between our tests and the numerical results from G09 is excellent. In all cases and at all temperatures, the curves closely track each other, in most cases leading to curves that are indistinguishable.  Although the abundances of several species change by many orders of magnitude throughout the runs, the two methods track each other within to 10\% in all cases except for
$H_2$ at $10^4$K, which is unimportant as a coolant but nevertheless consistent within a factor of $1.5$ at all times.
Furthermore, this agreement between methods is also seen for deuterium species, which are not shown in this figure as they follow H exactly, maintaining a $\frac{1}{6000}$ ratio between both species at all times.

At $T = 100 \rm \ K$, all ionized species quickly recombine with the free electrons to form neutral atoms. However, even during this relatively quick transition from ionized to neutral, H$^+$ and H$^-$ ions (not shown) persist for long enough to catalyze the formation of substantial amounts of molecular gas, leading to final H$_2$ molar fractions of $\approx 10^{-4}.$
At $T = 1000 \rm \ K$, the evolution is very similar to the $T = 100 \ \rm K$ case, although the species do not reach equilibrium as quickly, leading to even higher levels of H$_2$ formation. Finally, at $T = 10^4 \rm \ K$, it takes even longer for the ionized hydrogen to recombine, but in this case, less  molecular species are formed, 
as collisional dissociation of H$_2$ and HD are more prevalent, limiting the maximum amount of these species. 

Also apparent in these plot is the dependence of the species evolution on the density of the gas. Chemical reactions are fundamentally collisional processes whose rates are quadratic in number density.   Thus, as we are not considering three-body interactions, the timescale associated with chemistry should decrease linearly with the density.   This is seen for all temperatures and species shown in Figure~\ref{ctest}, as in every case each line is separated from its neighbor by a factor of $10$ in time, exactly corresponding to the density shift between cases.

\subsubsection{Effect of the Background Radiation}

\begin{figure}
\centering
\includegraphics[trim = 0mm 0mm 0mm 0mm,clip,scale=0.45]{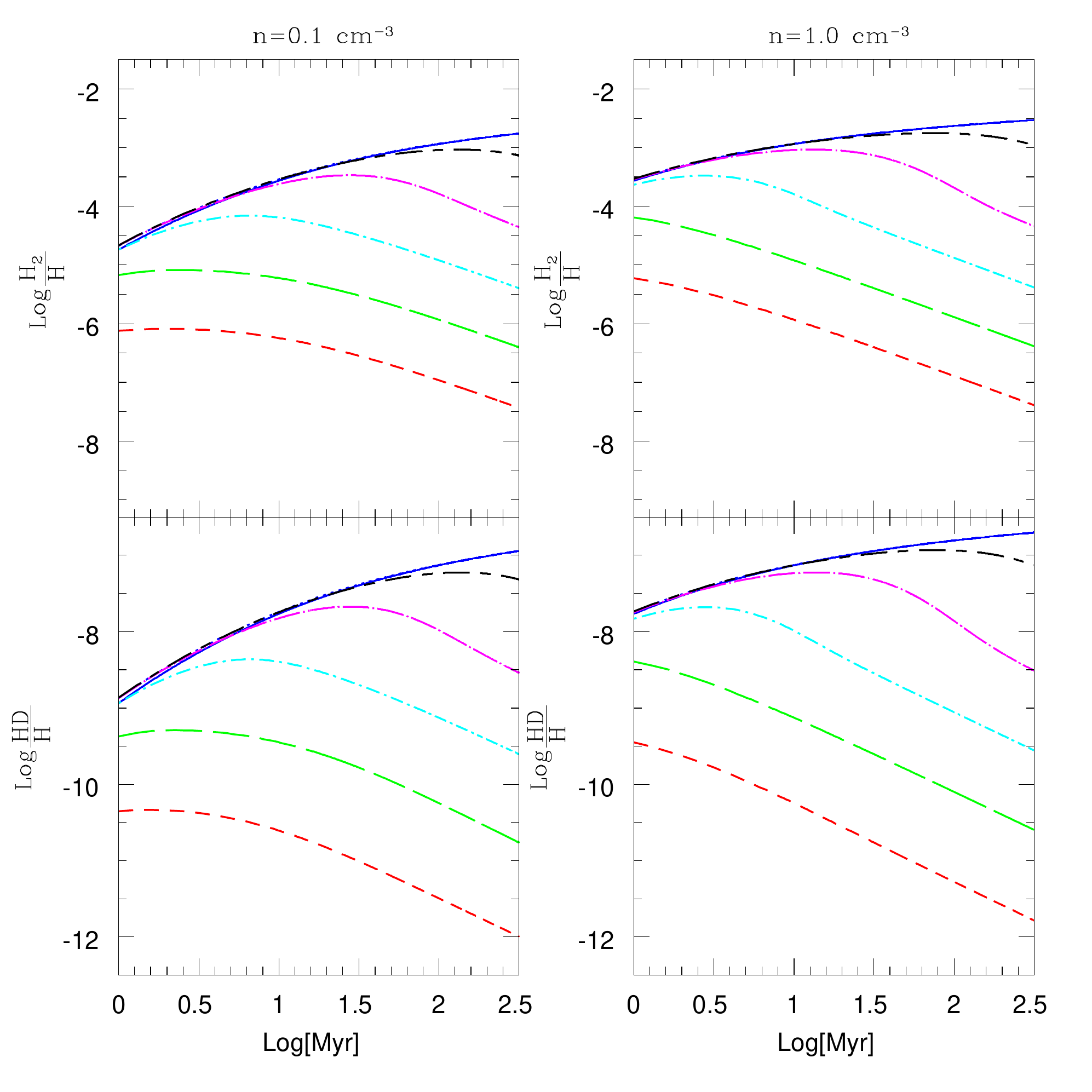}
\caption{Comparison of different UV backgrounds. The dotted line is the comparison from G09, the solid line is $J_{21}$ = 0, the short-and-long-dashed line is $J_{21}$ = $10^{-4}$, the dot-long-dahsed line is $J_{21}$ = $10^{-3}$, the dot-short-dashed line is $J_{21}$ = $10^{-2}$, the long-dashed line is $J_{21}$ = $10^{-1}$, and the short-dashed line is $J_{21}$ = 1.0. Time is given on the $x$-axis and number density of each species normalized by the number density of neutral hydrogen is given on the $y$-axis. Note that the solid line and dotted lines coincide with each other, demonstrating that we recover the expected results in the background-free case.}
\label{j21test}
\end{figure}

Background radiation with photon energies between 11.2 and 13.6 eV can excite and dissociate molecular hydrogen. In the absence of other coolants, this can have drastic effect on the evolution of the cloud. Two extremes are immediately apparent, a strong background case in which any H$_2$ or HD formed is quickly dissociated, and a background-free case in which no molecules are photodissociated. A simple test was constructed to study the effect of the background and determine a fiducial value for $J_{21}$. $J_{21}$ was varied between 0 and 1 at five different values. For each value of $J_{21}$ the number density was varied between $n$ = $10^{-1}$ and $1.0$ cm$^{-3}$. Each test was run at a constant temperature and constant density with evolving chemistry and no cooling. The results are given in Figure~\ref{j21test}. From this, we determine that only background levels at or above $J_{21}$ = $0.1$ give an appreciable difference in the abundance of H$_2$ and HD over a megayear timescale, which as we shall see below, is the timescale of shock-minihalo interactions.   At the same time, $J_{21}$ = 0.1 provides a reasonable upper limit to the level of background expected before reionization (\eg Ciardi \& Ferrara 2005).  
Therefore, we use this value as a fiducial value in the simulations with a background.

\subsection{Cooling}

The second major process added to the code was radiative cooling, which was divided into two temperature regimes. 
At temperatures $\geq 10^4$K, cooling results mostly from atomic lines of H and He, with bremmstrahlung radiation also becoming important at
temperatures above $10^7 K$.  Below $10^4$ K, on the other hand, the net cooling rate is determined by molecular line cooling from H$_2$ and HD, which, as it is an asymmetric molecule,
can radiate much more efficiently than H$_2$,  and thus can be almost as important although it is much less abundant.
Cooling from H$_2$ operates down to $T \leq 200$ K and to number densities $n > 10^4$ cm$^{-3}$
(Glover \& Abel 2008; Galli \& Palla; 1998), while HD which can cool the gas  to slightly lower temperatures and to
higher number densities (Bromm, Coppi, \& Larson 2002). 
As we are restricting ourselves to primordial gas in this study at
any given temperature the overall cooling rate, $\Lambda_{\rm Total}$, is the combination from both regimes,
\begin{equation}
\Lambda_{\rm Total} = \Lambda_{\rm Atomic} + \Lambda_{\rm Molecular}.
\end{equation}

Each cooling rate has the form:
\begin{equation}
\Lambda_{i,j} = n_{i} n_{j} \lambda_{i,j},
\end {equation}
where $\Lambda_{i,j}$ is the energy loss per volume due to species i and j, $n_{i}$ and $n_{j}$ are the number densities of each species, and $\lambda_{i,j}$ is the cooling rate in ergs $\rm cm^{-3} \ s^{-1}$. Cooling rates for the collisional excitation between H$_2$ and H, H$_2$, H$^+$, and e$^-$ and between H$_2^+$ and H or e$^-$ are taken from GA08. The cooling rate for the collisional excitation between HD and H is taken from Lipovka, N\'{u}\~{n}ez-L\'{op}ez, $\&$ Avila-Reese (2005). Finally, cooling rates from Hydrogen and Helium atomic lines are calculated using CLOUDY (Ferland, G.J., et al 1998). In calculating these rates, we followed the procedure described in  Smith \etal (2008) and used the ``coronal equilibrium" command which considers only collisional ionization.  The cooling curve was calculated assuming case B recombination for the recombination lines of hydrogen and helium, as discussed further in $\S 3.1$.

Any cooling routine contains a natural timescale that relates the total internal energy to the energy loss per time:
\begin{equation}
\tau_{\rm cool} = \frac{\alpha_{\rm cool} \times E_i}{\dot{s}}, 
\label{tau_cool}
\end{equation}
where $\alpha_{\rm cool}$ is a constant between 0 and 1, in all cases set at 0.1, $E_i$ is the internal energy, and $\dot{s}$ is the energy loss per time. 
Cooling rates are very dependent on temperature and species abundances and these quantities can change rapidly over a single chemical time step. 

\begin{figure}
\centering
\includegraphics[trim = 0mm 0mm 0mm 10mm,clip,scale=0.35,angle=90]{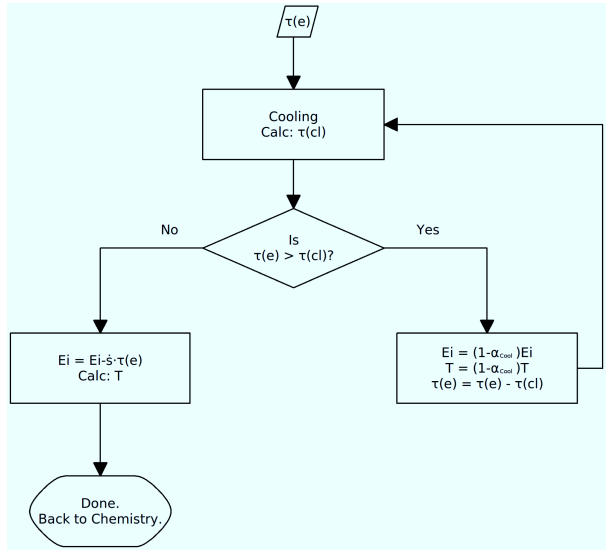}
\caption{Schematic view of the cooling subcycle. The time over which chemistry evolves $\tau(e)$ is used as the initial time step. This is compared against $\tau(cl)$ the cooling time step, as given by eq.~(\ref{tau_cool}). If the cooling time step is shorter than the evolved time step, then a portion of the internal energy and temperature are subtracted and the evolved time step is updated. The cooling time step is then recalculated. If the cooling time step is longer than the evolving time step then the internal energy is directly updated and used to calculate the new temperature. Once this is done, cooling is complete and we return to the chemistry routine}
\label{flow_cool}
\end{figure}

A method of subcycling over cooling time steps was developed to ensure that the correct cooling rates are used. An initial cooling time scale is calculated assuming $\alpha_{\rm cool}$ = 0.1 using eqn.~(\ref{tau_cool}) which is then compared to the chemical time step. If $\tau_{\rm cool}$ is smaller than the fraction of the chemistry time step then that fraction of energy is subtracted from the internal energy and temperature. The cooling rate and cooling time step is recalculated with the updated temperature. This continues until the chemistry time step is reached. This is schematically given in Fig.~\ref{flow_cool}.

\subsubsection{Cooling Tests}

As a test of our cooling routines, we reproduced the example curves given in Prieto \etal (2008). In this work, the authors present the effects of H$_2$ and HD cooling in a primordial gas.  The gas begins at an initial temperature of $T = 500$ K with initial number densities, relative to hydrogen: $n_{{\rm H}^+} = 10^{-4}; n_{{\rm H}^-} = n_{{\rm H}_2^+} = 10^{-12}; n_{{\rm H}_2} = 10^{-3}; n_{\rm D} = 10^{-5};  n_{{\rm D}^+} = 10^{-9}; n_{{\rm HD}} = 10^{-6}; n_{{\rm HD}^+} = 10^{-18}; n_{{\rm He}^+} = n_{{\rm He}^{++}} = 0.0$l and with initial hydrogen and helium densities of $\rho_{\rm H} = 0.75 \times \rho_{tot}$ and $\rho_{{\rm He}} = 0.24 \times \rho_{tot}$, where $\rho_{tot}$ is the total baryonic matter density.
	
\begin{figure}	
\begin{centering}
\includegraphics[scale=0.40]{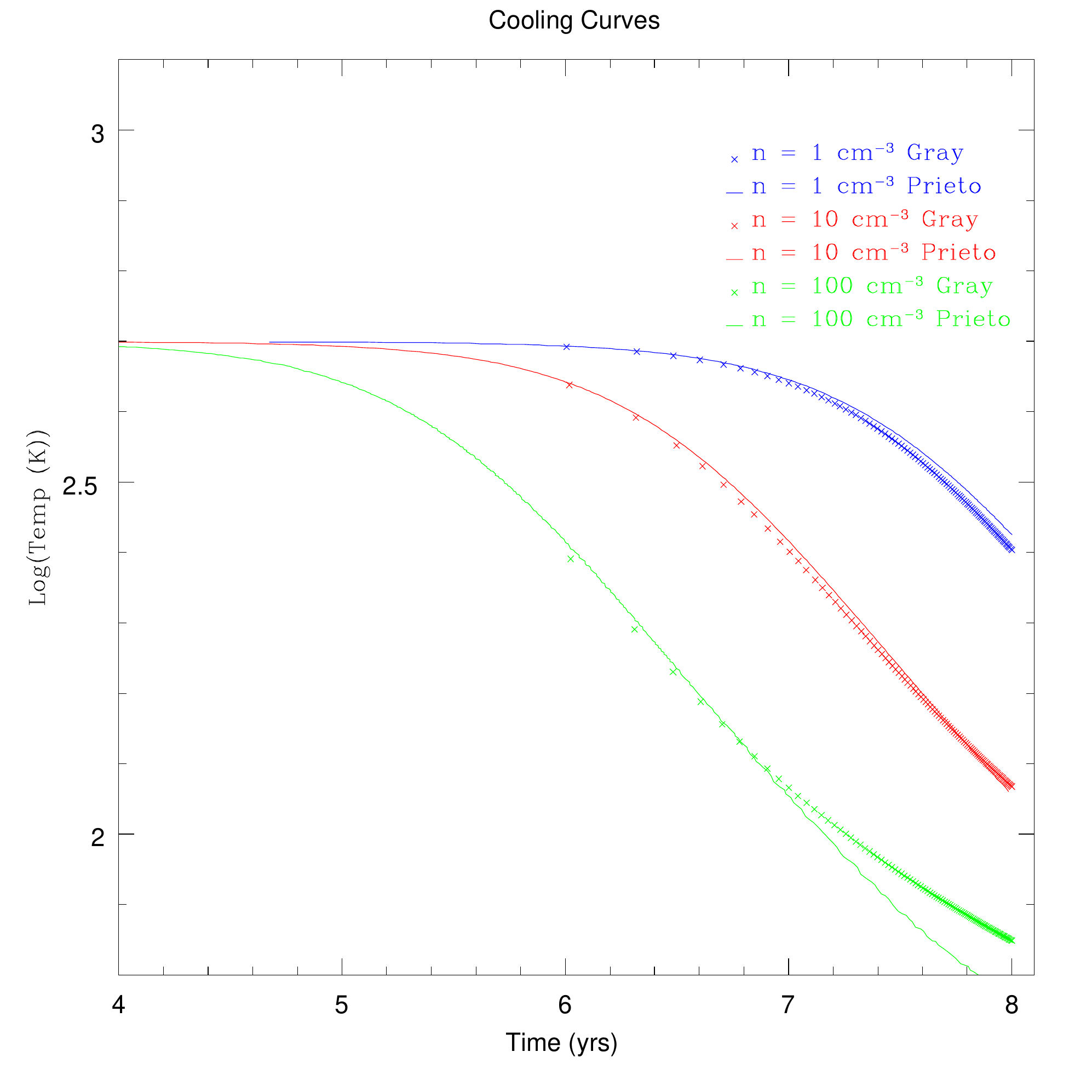}
\caption{Cooling tests. The solid lines are taken from Prieto \etal (2008) and compared to our model. The blue curves correspond to a number density $n=1.0 \rm \ cm^{-3}$, red to $n=10.0 \rm \ cm^{-3}$, and green to $n = 100.0 \rm \ cm^{-3}$. The temperature is not allowed to go below 50 K.}
\label{cool}
\end{centering}
\end{figure}

Three models were run with total number densities of $n_{tot} = 1.0, 10.0,$ and $100.0$  cm$^{-3}$.  Cooling was tracked for $10^8 \ \rm yrs$ with chemistry evolving simultaneously. The results of this calculation are shown in Figure~\ref{cool}, which indicates good agreement with Prieto \etal (2008). It should be noted that temperature evolution in this plot has a linear dependence of the number density of the gas. For example, a gas with ten times the number densities of another gas will cool ten times quicker.   This is again because most of the cooling is coming from the collisions between two species, in this case H$_2$ or HD and H.  
	
\begin{figure}
\begin{centering}
\includegraphics[scale=0.40]{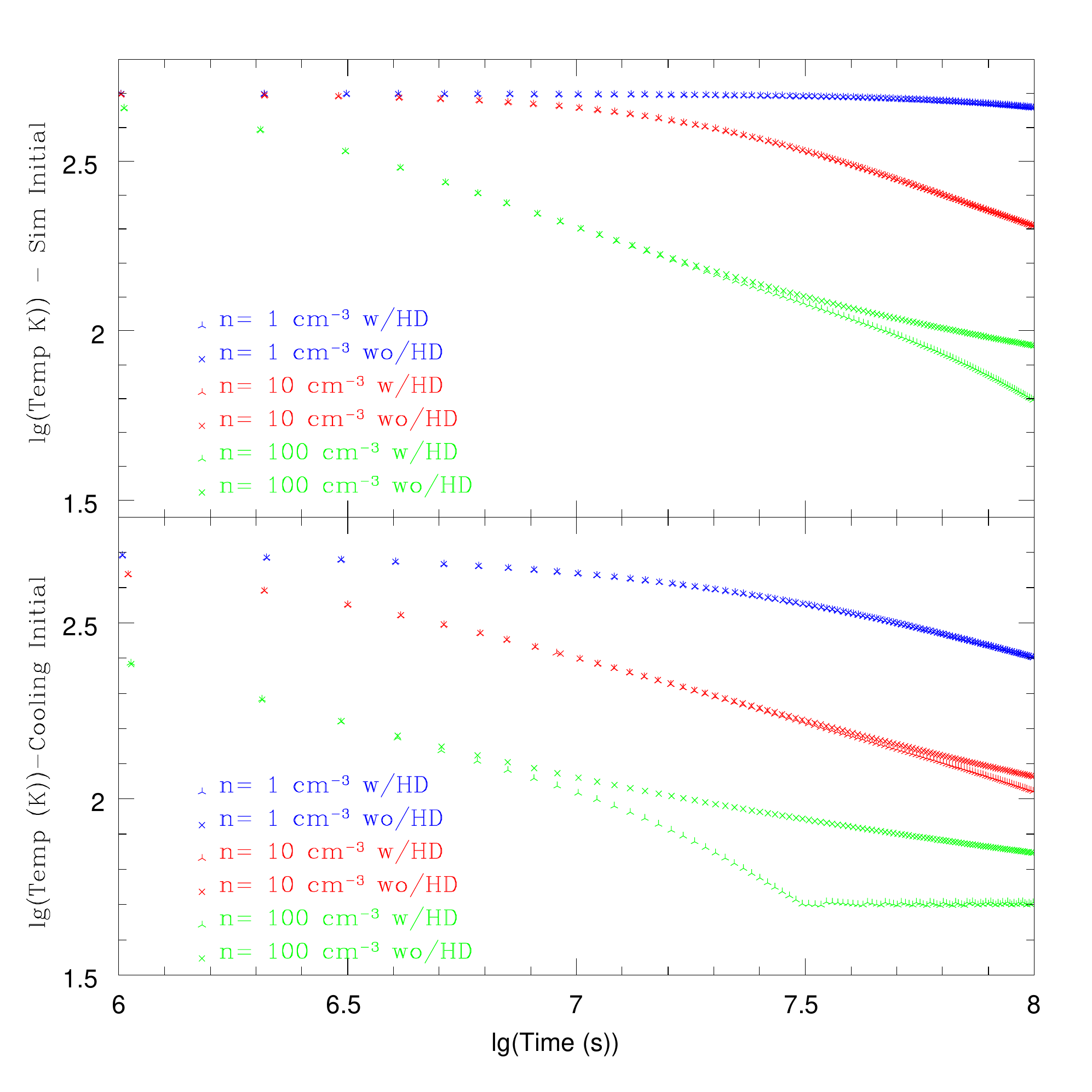}
\caption{Test of the impact of HD cooling. The top panel shows the results using primordial abundances while the bottom shows for abundances from the cooling test. Initially, the temperature is started at 500 K and evolved for 100 Myrs.}
\label{hdtest}
\end{centering}
\end{figure}

As mentioned above, HD can be more important than H$_2$ for gas cooling at higher densities and colder temperatures. To determine whether or not HD cooling is important in this simulation, we apply the cooling test to two different scenarios. First, we use the same initial abundances as described above and second, using primordial abundances with a small fraction ($0.01\%$) of each atomic species ionized. Each test was run twice, once with deuterium and once without. The results of these tests is given in Figure \ref{hdtest}. At high number densities, HD cooling does not have a perceivable effect. At intermediate temperatures, HD cooling is important for a gas with the initial abundances from the cooling tests Finally, at low temperatures, HD cooling is very important in both cases.

\section{Model Framework}

Having developed and tested the chemistry and cooling routines necessary to study minihalo-shock interactions, we next turn to the detailed shock-minihalo interactions.
Here we restrict our attention to a Cold Dark Matter (CDM) cosmology, with parameters are $h = 0.7$, $\Omega_0 = 0.3$ , $\Omega_{\Lambda} = 0.7$, and $\Omega_b = 0.045$ (\eg Spergel \etal 2007),  where $h$ is the Hubble constant in units of 100 km s$^{-1}$ Mpc$^{-1}$,  $\Omega_0 , \Omega_{\Lambda} ,$and $\ \Omega_b$ are the total matter, vacuum, and baryonic densities, respectively, in units of the critical density. For our choice of $h$, the critical density is $\rho_{crit}  = 9.2 \times 10^{-30}$ g/cm$^3$. 

\subsection{The Minihalo}

A simple model is used for the gas and dark matter of the protocluster whose collapse redshift of $z_c = 10$  (a cosmic age of $\approx \ 0.5$ Gyr) is taken to be just before the epoch of reionization, and whose total mass of $M_{c}  =  3.0 \times 10^6 M_{\odot}$ is taken to be on the large end of minihalos formed at this redshift.  The gas is assumed to have a primordial composition of 76\% neutral atomic hydrogen and 24\% neutral atomic helium by mass. 
Initially, the cluster has a mean density that is enhanced by a factor $\Delta = {\rm 178}$ (e.g. Eke, Navarro, \& Frenk 1998) above the background,  $\rho_c = \Delta \Omega_0 (1+z_c)^3 \rho_{\rm crit} = 6.54 \times 10^{-25} {\rm gm/cm^{3}} $. In this case, the cloud's virial radius is $R_c = 0.393\ \rm kpc$ and  virial velocity of $v_c = 6.55\ \rm {km/s}.$
We assume that the radial profile is given by Navarro \etal (1997)
\be
\rho(R) = \frac{\Omega_0\rho_c}{cx(1+cx)^2} \frac{c^2}{3F(c)} \  \rm gm/cm^3,
\label{rhor_eqn}
\ee
where $c$ is the halo concentration factor, $x$ = $R/R_c$ , and $F(t) \equiv \ln(1+t)-\frac{t}{1+t}$. We assume that as the gas collapses inside the dark matter halo, it is shock-heated to its virial temperature, $T_c = 1650\ \rm K$ and develops a density distribution of isothermal matter in the CDM potential well:
\begin{equation}
\rho_{\rm gas}(R) = \rho_0e^{ - \left( \frac{ v_{\rm esc}^2(0)-v_{\rm esc}^2(R) }{ v_c^2 } \right) } \rm gm/cm^3.
\label{rhogeqn}
\end{equation}
where the escape velocity is as a function of radius is given by $v(x R_{\rm vir}) = 2 v_c^2 [F(cx) + cx (1+cx)^{-1} ]  [x F(c)]^{-1}.$
From Madau \etal (2001), we take a typical value of the halo concentration to be $c = 4.8,$ although some observations suggest that high-redshift haloes maybe less concentrated than expected from this estimate (Bullock \etal 2001).  With this value of $c$, we can compute the central density as:
\begin{align}
\rho_0 & = \frac{(178/3)c^3\Omega_b e^A\left(1+z\right)^3}{ \int_0^c \left(1+t\right)^{A/t} t^2 dt} \rho_{\rm crit} & \nonumber \\ & = 39215 \  \Omega_b \rho_{\rm crit} \left(1+z_c\right)^3 & \nonumber \\ & = 2.16\times 10^{-23} {\rm gm/cm^{3}},
\end{align}
where $A \equiv 2c/F(c) = 10.3$ and $t=cx$.

To determine which case to use in our chemistry routine the optical depth for H$^+$, He$^+$, and He$^{++}$ recombination was calculated from this profile:
\begin{equation}
\tau_{\nu}(r) = \int^{r}_{r_0} \sigma_{\nu} n(r') dr',
\end{equation}
where $\sigma_{\nu}$ is the cross section of interaction and $n(r')$ is the number density. 
For hydrogen-like atoms the cross section is
\begin{equation}
\sigma_{\nu} = \frac{7.91 \times 10^{-18}}{Z^2} \left(\frac{\nu_1}{\nu}\right)^3 \ {\rm g} \rm \ \ cm^2,
\label{sig}
\end{equation}
where $h\nu_1 = 13.6 \ Z^2 \rm \  eV$ and $Z$ is the nuclear charge. Calculating the optical depth from eqs.~(\ref{rhogeqn} - \ref{sig}) yields the results given in Table 1. We assume for the case of He$^{++}$ recombination, that the surrounding helium is singly ionized.  In all cases the optical depth is much greater than 1 (see \S 3.0) and therefore we use case B rates for all recombination reactions in our fiducial simulations. 

\vspace{5mm}
\begin{table}[ht]
\centering
\caption{Optical Depths. $\tau$ is the optical depth to the center of the cloud. In all cases the optical depth is much greater than 1.} 
	\begin{tabular}{| c | c | c |}
	\hline
	$\rm Species$&$\rm Incident \  Energy  \ eV$&$\rm \  \tau $\\
	\hline
	$\rm H^+$&$13.6$&$3553.9$ \\
	$\rm He^+$&$24.6$&$1049.2$ \\
	$\rm He^{++}$&$54.4$&$280.6$ \\
	\hline
	\end{tabular}
\end{table}
\vspace{5mm}

 \subsection{Gravity}

As the minihalo in our simulation is made up of both gas and collisionless dark matter, 
a two-part gravity scheme was required. First, the Ricker (2008) multigrid Poisson solver was used to calculate the gravitational potential due to the gas component.
Second, the acceleration due to the total matter was calculated using Eq.\  (\ref{rhogeqn}) above. The general equation for the gravitational acceleration of an ideal gas is given by 
$a_{\rm Grav} = \frac{k}{\rho(R) m_p} \left[ T(R) \frac{\partial \rho(R)}{\partial R} + \rho(R) \frac{\partial T(R)}{\partial R} \right] \  \rm cm/s^2.$
To account for the dark matter halo's contribution to gravity, we calculated the gravitational acceleration of the total matter and 
subtracted its contribution from the baryonic matter in the initial configuration
using the above equation with constant temperature. Finally, we added the gravitational contribution from the self-gravity of the gas.  The total acceleration is given simply by:
\begin{equation}
a_{\rm Tot}  = a_{\rm M,0} - a_{\rm gas,0} + a_{SG},
\end{equation}
where $a_{M}$ is the acceleration from the total initial mass density as given by Eq.\ (\ref{rhogeqn}), $a_{\rm gas,0}$ is the contribution from the baryonic matter in the initial configuration, and $a_{SG}$ is the contribution from the self-gravity calculated from the Poisson solver. Initially, when the minihalo gas is in hydrostatic balance with its surroundings, these last two terms will cancel each other and the cloud will remain unchanged. When the cloud is disrupted and cooling takes effect, the self-gravity will cause the cloud to collapse.

The above equations are correct up to the viral radius of the cloud. To ensure a smooth density transition from the cloud, we simply keep the gas outside of the cloud be gravitationally bound to the cloud and solve for the expected density profile. The acceleration here is then,
\begin{equation}
a_{\rm Grav} =-\frac{GM(R > R_c)}{R^2},
\end{equation}
and the density is
\begin{equation}
\rho(R > R_c) = \rho(R_c)e^{\left({\frac{R_0}{R} - \frac{R_0}{R_c}}\right)},
\end{equation}
with $R_0$ = $G M_c m_p$/$k_b T$. Here $G$ is the gravitational constant,$m_p$ is the mass of a proton, $k_b$ is Boltzmann's constant, and, as above $M_c$ and $T_c$ are the mass and temperature of the cloud. As $R \rightarrow \infty$, the density goes down to a small fraction of $\rho(R_c)$.  A test of our gravity routine showed that the cloud was able to maintain hydrostatic balance for many dynamical times in the absence of an impinging galaxy outflow.

\subsection{The Outflow}

A Sedov-Taylor solution is used to estimate the properties of the galactic outflow. The initial input energy is taken to be $E = \epsilon E_{55} ({\rm ergs}),$ 
where $E_{55}$ is the energy of the supernovae driving the wind in units of $10^{55}$ ergs,  and the wind efficiency $\epsilon$ is derived from the amount of kinetic energy from the supernovae that is channeled into the outflow. The shock expands into a gas that is $\delta$ times greater than the background at a redshift of $z_c$. 
As in Scannapieco \etal (2004),  we assume that the cloud is a distance $R_s =  3.6 \rm \ kpc$, using fiducial values: $z_c = 10$, $E_{55} = \rm 10$, $M_6 = 3$, $\delta=44$ and $\epsilon = 0.3$

With these values, the velocity of the blast front is $v_s = 415 \ \rm km\ s^{-1},$ when it reaches the minihalo, and
the resulting temperature of the fully-ionized post-shock medium is $T = 2.4 \times 10^6$ K.
By the time the shock has covered the separation distance, $R_s$, it will have entrained a mass
\be
M_{s, \rm Total} =  4.4 \times 10^{7} \rm \ M_{\sun},
\ee
with a surface density of
\be
\sigma_{s} = 2.6 \times 10^{5} \rm \ M_{\sun} \ kpc^{-2}.
\ee
Note that while the above equations are the solutions that come from a simple spherical blast wave, the wind in the simulation is still well-approximated by a plane wave solution, because the size of cloud is much smaller than the distance between the supernova and the cloud. 

To model this wind,  a time-dependent boundary condition is imposed at the leftmost boundary of the simulation volume. The expected lifetime of the shock is given by $\sigma_s = v_{\rm post} \, \rho_{\rm post} \, t_s,$ where $v_{\rm post} $ is the post-shock velocity of the blast wave, $\rho_{\rm post}$ is the post-shock density, and $\sigma_s$ is the surface density of the entrained material.  Solving for $t_s$ and putting in the appropriate values, the expected shock life time is $t_s = 2.5 \rm \ Myr$. After this time the shock begins to taper off with the density decreasing and temperature increasing and keeping  the pressure constant. This is done to prevent the excessive refinement that a sharp cutoff would cause. The density falls off as 
\begin{equation}
\frac{\rho(t)}{\rho(0)}  = 0.01 + 0.99 e^{-\tau_s/1.5}
\end{equation}
and the temperature rises as
\begin{equation}
\frac{T(0)}{T(t)}  = 0.01 + 0.99 e^{-\tau_s/1.5} \ ,
\end{equation}
where $\tau_s$ is defined as $\frac{t - 1 \Myr}{1 \Myr}$. This also prevents the hydrodynamic (Courant) timestep from becoming extremely short
behind the shock, in order to maintain pressure equilibrium in an extremely rarified medium.

Note that in this initial study, the dark matter and gas distributions have been somewhat idealized, and more complicated geometries could be used to model these components in greater detail. For example, a triaxial instead of a spherical distribution could be assumed for the dark matter halo, inhomogeneities could be added to the minihalo gas, and the shock could be assumed to impact the minihalo off-axis.  While each of these possibilities would be qualitatively interesting, and naturally alter the final outcome of the halo, they are nevertheless beyond the scope of this study.

\section{Results}

	Our simulations were carried out in a rectangular box with an effective volume of 3.2 $\times 10^9$ $\rm pc^3$. The $y$-axis and $z$-axis were the same length of 1170 pc and range between [-585, 585] pc while the $x$-axis was twice as long, stretching between [-585, 1170] pc.  The shock started on the left boundary while the cloud was centered at [0,0,0] pc. As hydrodynamic refinement criteria, FLASH uses the second derivative of  ``refinement variables," normalized by their average gradient over a cell. If this was greater than 0.8, the cell was marked for refinement, and if all the cells in a region lie below 0.2, those cells were marked for derefinement. 
	
	A detailed summary of the runs performed is given in Table 2. The runs are labeled as either high or low resolution (H or L), whether atomic H-He recombination follows Case A or Case B (A or B), and whether we impose a UV background (Y or N).  The high-resolution, Case B, no-background run (HBN) 
is taken to as our fiducial run and compared against other choices of parameters below.

\vspace{5mm}
 \begin{table}[ht]
\centering
\caption{Summary of the numerical simulations in this study.}
	\begin{tabular}{| c | c | c | c | c  |}
	\hline
	$\rm Name $&$\rm  l_{ref} $&$\rm Resolution (pc) $&$ \rm Cooling \  Mode $&$ \rm{Background \ (J_{21})} $\\
	\hline
	$\rm HBN $&$ 6 $&$ 4.55 $&$ \rm Case \  B $&$ \rm{ 0} $ \\
	$\rm LBN  $&$ 5 $&$ 9.11 $&$ \rm Case \  B $&$ \rm{  0} $ \\
	$\rm HBY $&$ 6 $&$ 4.55 $&$ \rm Case \  B $&$ \rm{ 10^{-1}} $ \\
	$\rm LBY  $&$ 5 $&$ 9.11 $&$ \rm Case \  B $&$ \rm{  10^{-1}} $ \\
	$\rm HAN $&$ 6 $&$ 4.55 $&$  \rm Case \ A $&$ \rm{  0} $ \\
	$\rm LAN  $&$ 5 $&$ 9.11 $&$ \rm Case \ A $&$ \rm{  0} $\\
	\hline
	\end{tabular}
\end{table}

\vspace{5mm}

\subsection{Hydrodynamic Evolution}

\begin{figure*}
\begin{centering}
\includegraphics[scale=0.75, trim = 10mm 5mm 10mm 10mm] {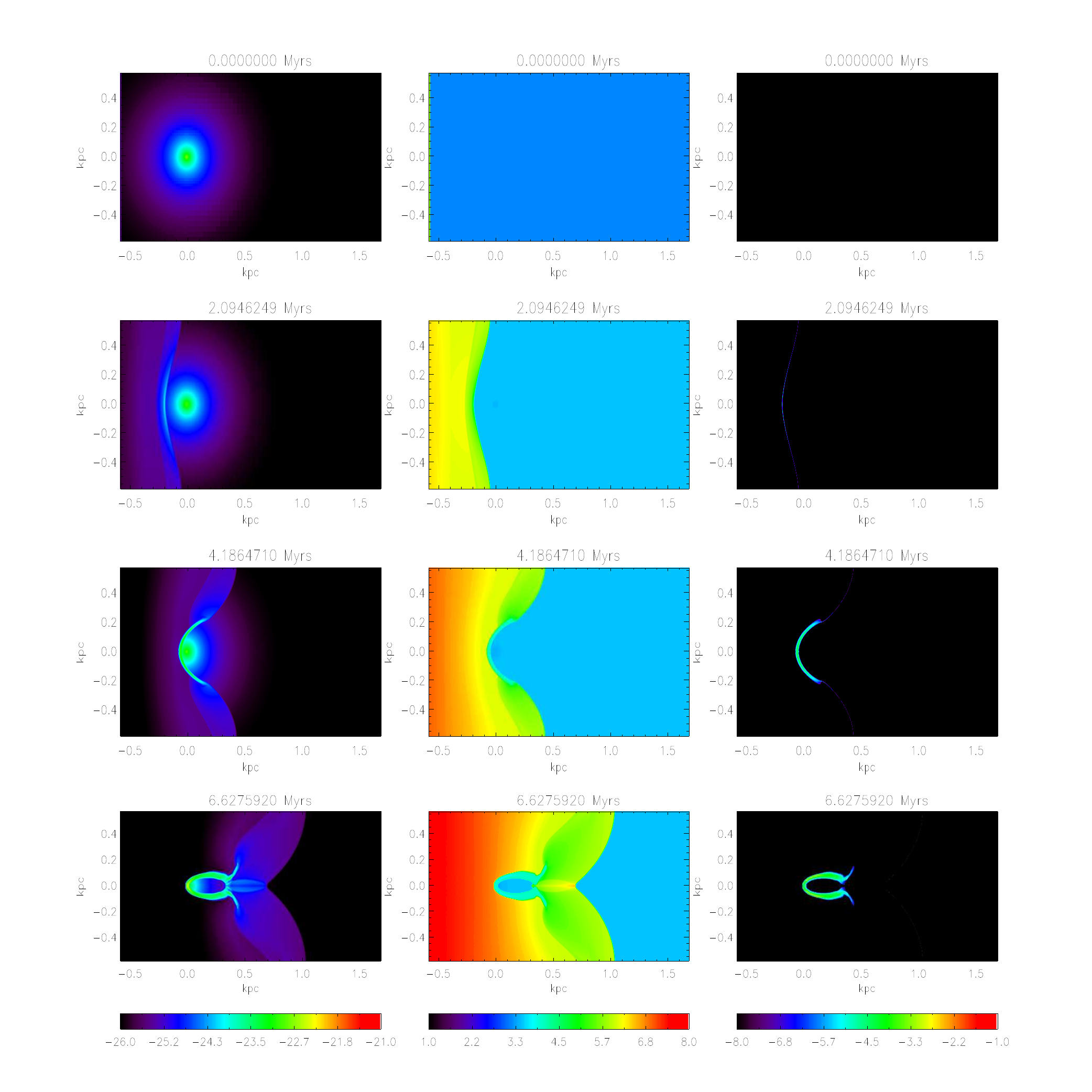}  
\caption{Initial evolution of the fiducial run, HBN, from $t=0$ through $t=t_{\rm ic}$ the time the shock completely surrounds the cloud. Each row shows the conditions in the central in a slice through the center of the simulation volume at times of 0 (top), 2.1 (second row), 4.2 (third row), and 6.6 Myrs (bottom row). The first column shows contours of the log of density from $\rho_{\rm gas} = 10^{-26}$ to 10$^{-21}$ g cm$^{-3}$, which corresponds to number densities from $n \approx 10^{-2}$ to $10^{2}$, the second column shows contours of the log of temperature from $T= 10$ to $10^8$K, and  the third column shows contours of the log of the H$_2$ mass fraction from $X_{\rm H_2} = 10^{-8}$ to $10^{-1}$. }
\label{sim1}
\end{centering}
\end{figure*}

\begin{figure*}
\begin{centering}
\includegraphics[scale=0.55, trim = 10mm 0mm 0mm 0mm]{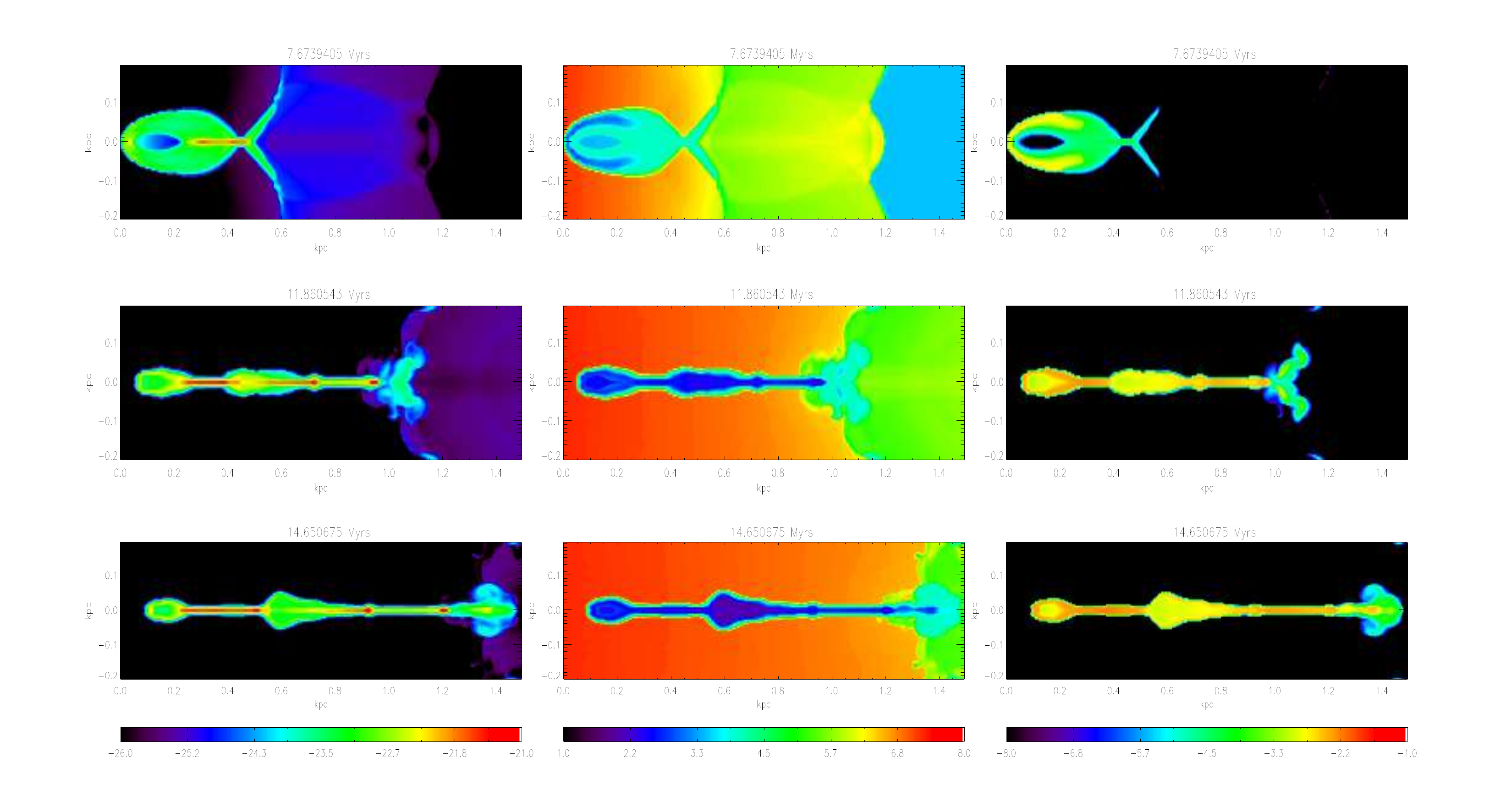} 
\caption{Final evolution of the cloud from the propagation of the reverse shock across the cloud at $t=7.7$ Myrs (top row), to collapse at $t=11.8$ Myrs (center row), through to the end of the simulation at $t=14.7$ Myrs (bottom row). The panels have been cropped to show only the extended mass along the $x$-axis. Columns, values, and contours are the same as Fig.~\ref{sim1}.}
\label{sim2}
\end{centering}
\end{figure*}

In this simulation, several distinct stages of evolution are identified during the interaction between the cloud and the outflow, as shown in Figures ~\ref{sim1} and ~\ref{sim2}. Initially, the cloud is in hydrostatic equilibrium as the shock enters the simulation domain. If it were not supported by pressure, the cloud would collapse on the free-fall time which, using the average cloud density, is
\begin{equation}
t_{\rm ff} = \sqrt{\frac{3 \pi}{32 G \rho}} \approx 100 \rm \ Myr.
\end{equation}
As the cloud is initially in hydrostatic balance, the initial sound crossing time is similar to the free-fall time. 

As the shock contacts and surrounds the cloud, it heats and  begins to  ionize the gas. The shock completely envelops the cloud on a characteristic ``intercloud" crossing time scale, defined by Klein \etal (1994) as
\begin{equation}
t_{\rm ic} = \frac{2 R_c}{v_{s}} \approx 4.5 \rm \ Myr.
\end{equation}
As the cloud is enveloped, the shock moves through the outer regions fastest and ionizes this gas first. This in turn promotes rapid molecule formation as the gas cools and recombines incompletely, leaving H$^+$ and H$^-$ to catalyze the formation of H$_2$ and HD.  Interestingly, because the shock slows down as it moves through denser material, the gas behind the center of the halo remains undisturbed until the enveloping shocks meet along the axis at the back of the halo.   The leads to a ``hollow" H$_2$ distribution at 6.6 Myrs, in which the molecular coolants are confined to a shell surrounding the undisturbed, purely atomic gas.

After the enveloping shocks collide at the back of the cloud, a strong reflected shock is formed that moves away from the rear of the cloud and back through the halo material. Without cooling, this reflected shock would eventually lead to cloud disruption (Klein \etal 1994). However in our case, the shock has the opposite effect.
It moves through the cloud, and the gas is briefly ionized, but then quickly cools and recombines, forming H$_2$ and HD throughout the cloud.
This can be seen in the upper row of Figure~\ref{sim2}, which shows the conditions at $\approx 8$ Myrs.

At this point the cloud is denser, smaller, and full of new coolants. Using the conditions from the center of the cloud 8 Myr after the start of the simulation, we calculate new timescales. Now the freefall time is $21$ Myr and the sound crossing time is $\approx 27$ Myr. The cloud is cold and dense enough to start collapsing.

The timescale for the formation of H$_2$, given in GA08, is 
\begin{equation}
t_{\rm H_2} = \frac{X_{H_2}}{k_1X_{e}n} \rm (s),
\end{equation} 
where $X_{{\rm H}_2}$ is the mass fraction of H$_2$, X$_e$ is the mass fraction of electrons, $k_1$ is the reaction rate for the formation of H$^-$ (H + e$^-$ $\rightarrow$ H$^-$ + $\gamma$), and $n$ is the total number density ($\approx$ 1 cm$^{-3}$ at 8 Myrs).  Initially, as the shock begins to impact the cloud, this timescale is very short, on the order of 0.1 Myrs to get a final abundance of $\approx$ $10^{-5}$. As the abundance of H$_2$ increases and the abundance of electrons decrease, this timescale quickly increases. Although as the cloud collapses the density increases which lowers this timescale.

The H$_2$ cooling timescale, given by Klein \etal (1994) is
\begin{equation}
t_{\rm cool} = \frac{1.5 n k T}{n_{{\rm H_2}} n_{{\rm H}} \Lambda_{{\rm H,H_2}}} \rm \ (s),
\end{equation}
where $n_{{\rm H}_2}$ and $n_{H}$ are the number densities of H$_2$ and H respectively, and $\Lambda_{{\rm H,H_2}}$ is the cooling rate between H and H$_2$. 
At 8 Myr the H$_2$ cooling time in most of the cloud is only $0.2$ Myr, meaning that pressure support drops dramatically after this time. Any expansion due to shock heating is halted as the gas is quickly cooled by H$_2$ and HD as they form.  Furthermore, as the cloud collapses, the chemistry and cooling timescales decrease, rapidly  accelerating the collapse.

The final state of the cloud in our simulation is a thin cylinder stretching from the center of the dark matter halo to several times the initial virial radius. The temperature of this gas is 100 to 200 degrees, much colder than the initial virial temperature. The gas is also much denser than the initial minihalo, reaching values of up to $10^{-21} \ \rm \ cm^{-3}$ or $n \approx 10^3$  cm$^{-3}$, in the center of the cloud, and even this density is probably only a lower limit set by the resolution of our simulation.  On the other hand, the cloud is quite extended along the $x$-axis, with substantial differences in velocity along the cylinder.   Thus it is continually stretched and fragments  until the end of the simulation at 14.7 Myrs (Row 3 in Figure \ref{sim2}). 

\begin{figure*}
\centering
\includegraphics[scale=0.4, trim= 0mm 0mm 0mm 0mm]{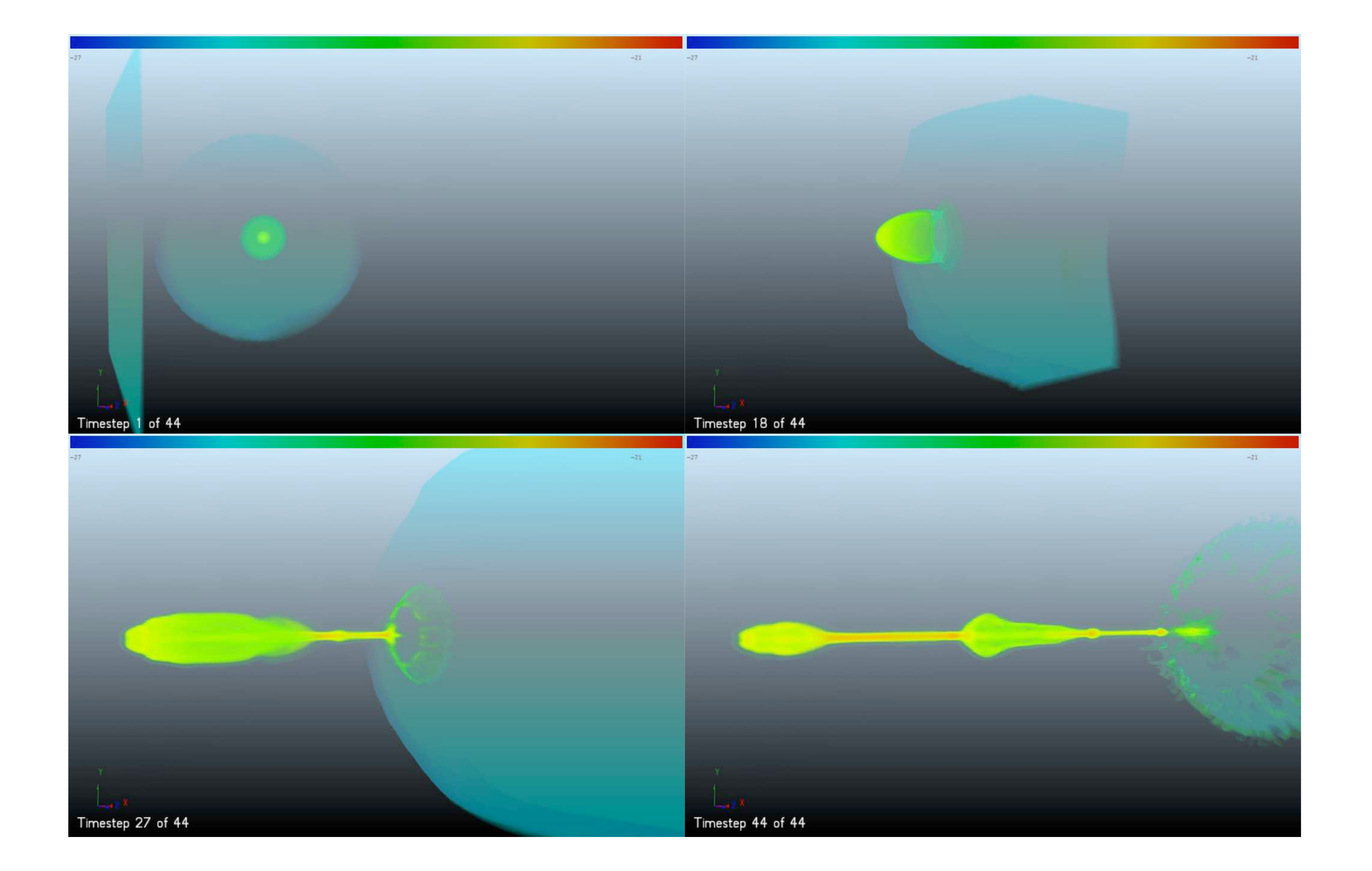}
\caption{Rendered density snapshots from the fiducial run, HBN. Colors show contours of log $\rho$ from $10^{-27}$ to $10^{-21}$ cm$^{-3}$.  The top left panel, $t =$ 0.0 Myrs, shows the initial setup with the stationary minihalo and the shock entering on the left. Top right, $t =$ 6.3 Myrs, shows the state of the cloud as the shock as it envelopes the cloud. Bottom left,  $t =$ 9.4 Myrs, shows the cloud during collapse and cooling. Finally, bottom right, $t =$ 14.7 Myrs, shows the final state of the cloud as it is stretched. Dense clumps can be seen in this panel, which we expect to become compact stellar clusters. }
\label{helios}
\end{figure*}

Figure~\ref{helios} shows  rendered density contours of the major stages of evolution of the cloud from $t=0$ through the end of the simulation.
The first panel shows the initial configuration, with the cloud in hydrostatic equilibrium, and the shock front entering from the left side of the simulation volume. The next panel shows the cloud after being impacted by the shock, highlighting the  density enhancement in the outer shell of the minihalo gas. The bottom left panel shows the cloud as it begins to cool and collapse, at a time at which the reverse shock has already passed though the cloud and coolants are found throughout the shocked, recombined material.    Finally, the last panel shows the distribution at the end of the simulation.  The cloud has now been stretched over a large distance and much of its mass has been accelerated to above the escape velocity, moving outside of the dark matter halo.  The dense knots of this material in this figure are tightly gravitationally bound,  have number densities approaching $10^{3}$  cm$^{-3},$  and are destined to form extremely compact stellar clusters.

\subsection{Stellar Clusters}

While the collision happens on order of the shock crossing time of the halo, the final distribution of the clumps evolves on the longer timescale
defined by $R_{\rm vir}/v_{c} \approx 100$ Myrs.
To study the final state of the stretched and collapsed distribution without continuing the simulation out to such extremely long times, 
we divided the $x$-axis into 100 evenly spaced bins between $x=0$ kpc and $x=1.4$ kpc. We then calculated the mass of each bin by summing up the gas from each cell from the FLASH simulation 
in a cylinder with a 24 pc radius and length of the bin. Similarly, we calculated the initial velocity of each bin  by adding the momentum from each cell within this cylinder and dividing by the total mass  in each bin. 

We evolved this distribution forward in time using a simple numerical model, which assumed that motions were purely along the $x$ axis and pressure was negligible at late times. In this case,  acceleration could be calculated directly from the gravity between each pair of  particles and from the potential of the dark matter halo.   Furthermore, if any given particle  moved past the  particle in front of it we merged them together, adding their masses and calculating  a new velocity from momentum conservation. 

\begin{figure}
\centering
\includegraphics[scale = 0.45, trim= 10mm 0mm 0mm 0mm]{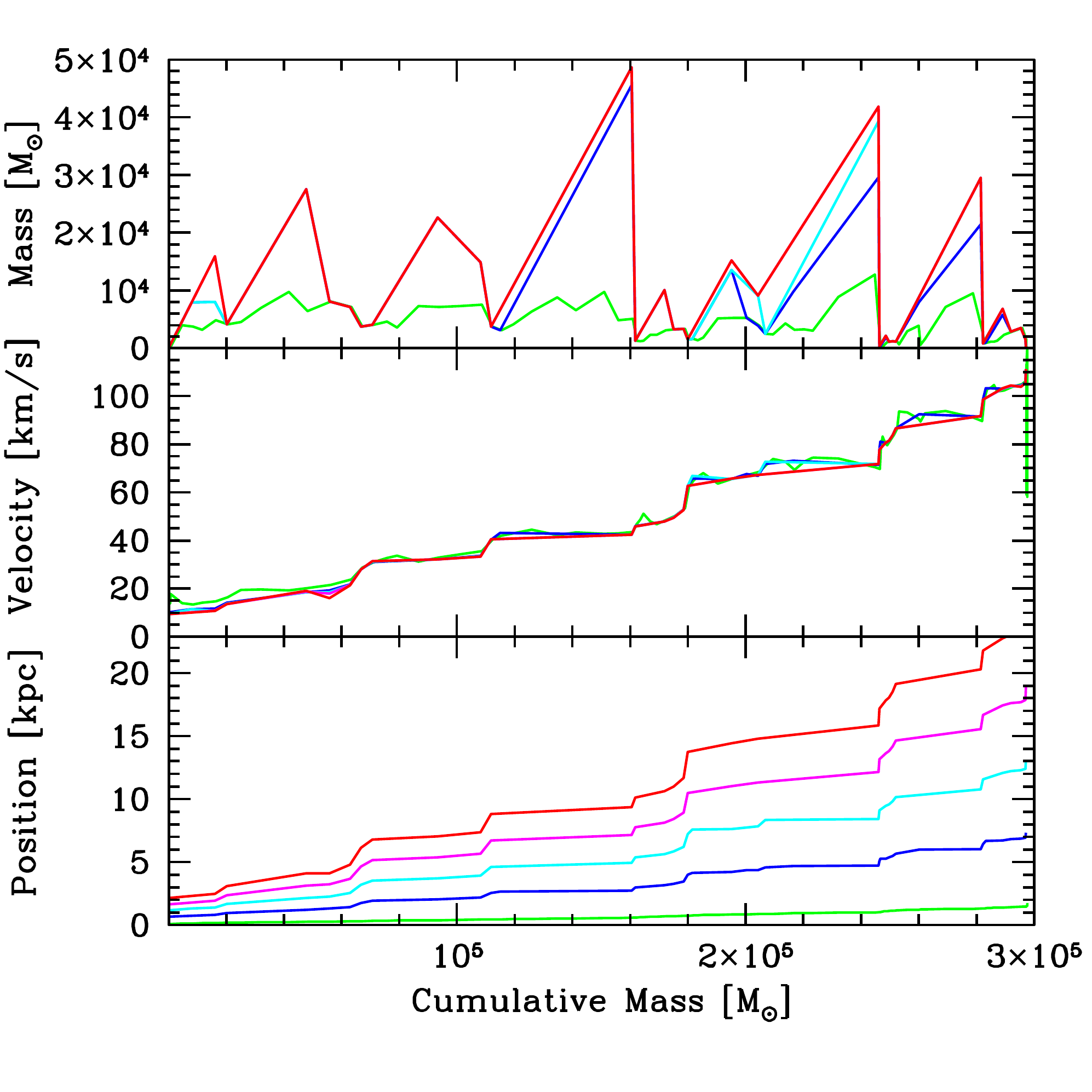}
\caption{Evolution of the cloud up to 200 Myrs after the end of the simulation. The $x$-axis in each panel is the cumulative mass in solar masses. The top panel shows the mass of each particle, the middle panel shows their velocities, and  the bottom panel shows their positions. The solid green lines show the profile at the end of the simulation $t_f = 14.7$ Myrs, the dotted blue lines show the profile 50 Myrs later, the short-dashed cyan lines show the profile at  $t_f + 100$ Myrs, the dot-short dashed  magenta lines show the profile at $t_f +150$ Myrs, and finally the short dash-long dashed red lines show the profile at $t_f + 200$ Myrs. As time progresses we find that much of the material in the linear feature from Fig.~\ref{sim2} merges together. Most of this merging is complete by 100 Myrs after the end of the simulation. }
\label{evolve}
\end{figure}

Evolving the distribution in this way for an additional 200 Myrs past the end of the simulation yielded the results shown in  Figure~\ref{evolve}. As the stretched cloud continues to move outward, particles begin to attract each other, and eventually merge together to create larger clumps.  By 100 Myrs most of the particles have merged, after which their motions are purely ballistic. This can seen in the top panel as the lines for the late times overlap each other and in the middle panel as the velocity profiles overlap.

At the final time of 200 Myrs after the end of numerical simulation,  three small, stable clumps with masses of 5.0$\times 10^4$ M$_{\odot},$ 4.0$\times10^4$ M$_{\odot},$ and $3 \times 10^4$ $M_\odot,$ as can be seen in the top panel of Fig.~\ref{evolve}.   Each of these new peaks is located far outside of the original dark matter halo.

\subsection{Case A vs. Case B}
\begin{figure*}
\centering
\includegraphics[scale=0.45, keepaspectratio=true, trim= 20mm 0mm 0mm 0mm]{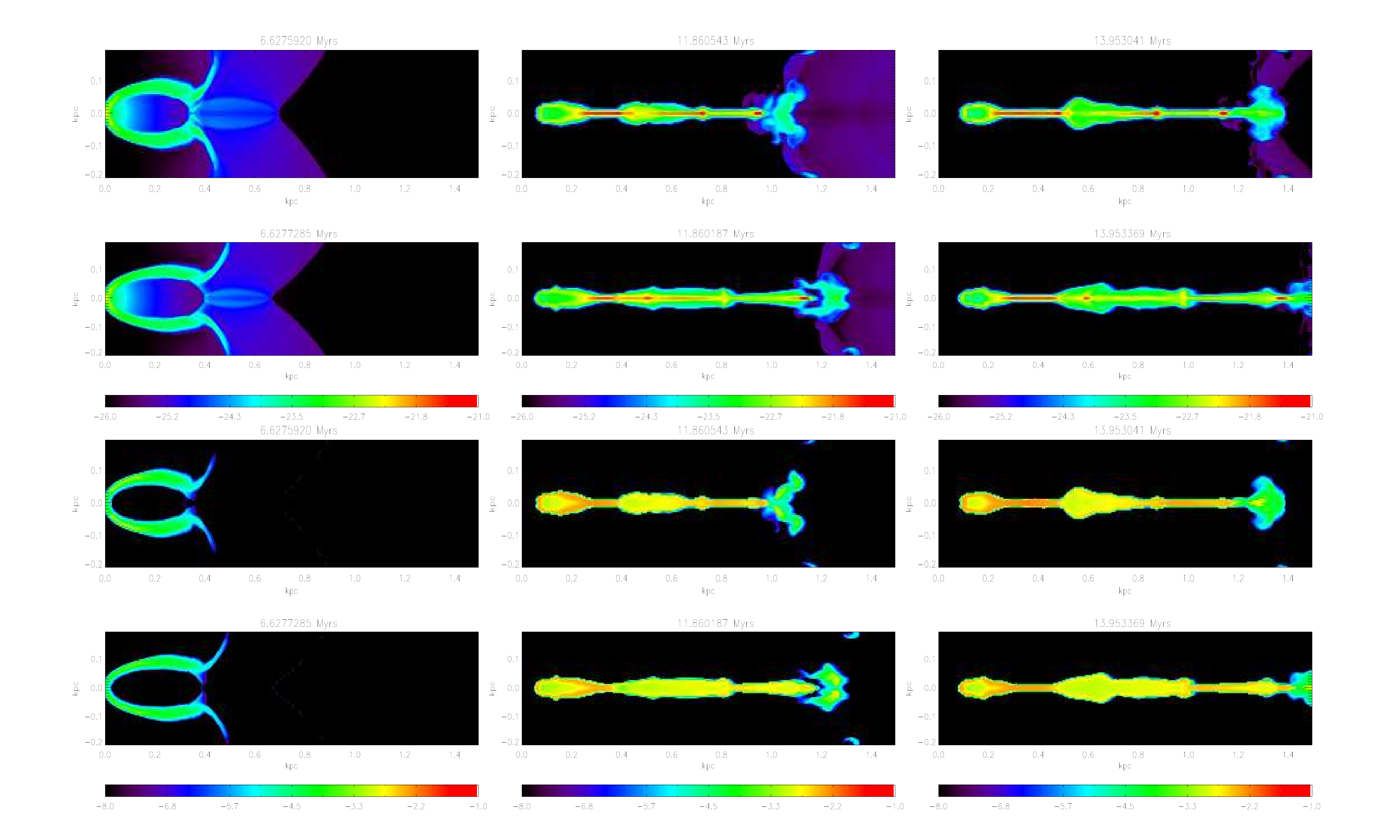}
\caption{Comparison between Case A and Case B cooling and chemistry rates.   In this plot
time varies across columns, moving from $t=6.6$ Myrs (left column), to $7.7$ Myrs (center column), to $14.0$ Myr (right column).  The upper two rows show the density in the central slice from the fiducial, case B run (HBN, top row), and the case A run (HAN, second row), with log contours ranging from $\rho = 10^{-26}$ to $10^{-21}$ g cm$^{-3}.$  The lower two rows show the H$_2$ mass fraction in the fiducial run (third row) and the Case A run (bottom row).  Here the log H$_2$ mass fraction contours range from $X_{\rm H_2} = 10^{-8}$ to $10^{-1}$.}
\label{compcases}
\end{figure*}

At temperatures above $10^{4}$ K, the primary source of cooling is atomic lines from hydrogen and helium. Although we have shown that for the primordial cloud, Case B rates should be used for both the chemical network and cooling functions, Figure~\ref{compcases} shows a comparison between our fiducial run, HBN, and a run in which reaction and cooling rates are taken for case A recombination (HAN). The high temperature Hydrogen-Helium cooling curve is taken from Weirsma \etal (2009).
The upper panels show density contours and the bottom show contours of H$_2$ abundance. 

As expected, the Case B simulation produces greater molecular coolant abundance at similar overall densities. This difference can be seen in the lower two panels of the first column of Fig.~\ref{compcases} at 6.63 Myrs.  To remain in pressure support as the cloud becomes denser from the shock, the cloud must get hotter. However, because Case A cools slightly faster, this support is quickly removed and the cloud takes on a more extended shape as evident in the first two columns of Fig.~\ref{compcases}.  Although, by 14 Myrs the abundance of molecular coolants are very similar between HBN and HAN with each containing $X_{H_2} \approx 10^{-2.5}$. 

In both cases, the fate of the minihalo gas is the same.  Atomic cooling occurs sufficiently rapidly to sap the shock of its energy and drop the post-shock temperature to $\approx 10^4$ K,  and nonequillibrium processes step in to provide molecular coolants below $10^4$ K. 
The gas is then able to collapse and form into a long dense filament within which clumps are formed.   In fact the only substantial differences between the runs are the details of the distributions of clumps, which is somewhat more extended in the case A run as compared to the case B run.

\subsection{UV Background}

\begin{figure*}
\centering
\includegraphics[scale=0.45, trim= 15mm 0mm 0mm 0mm]{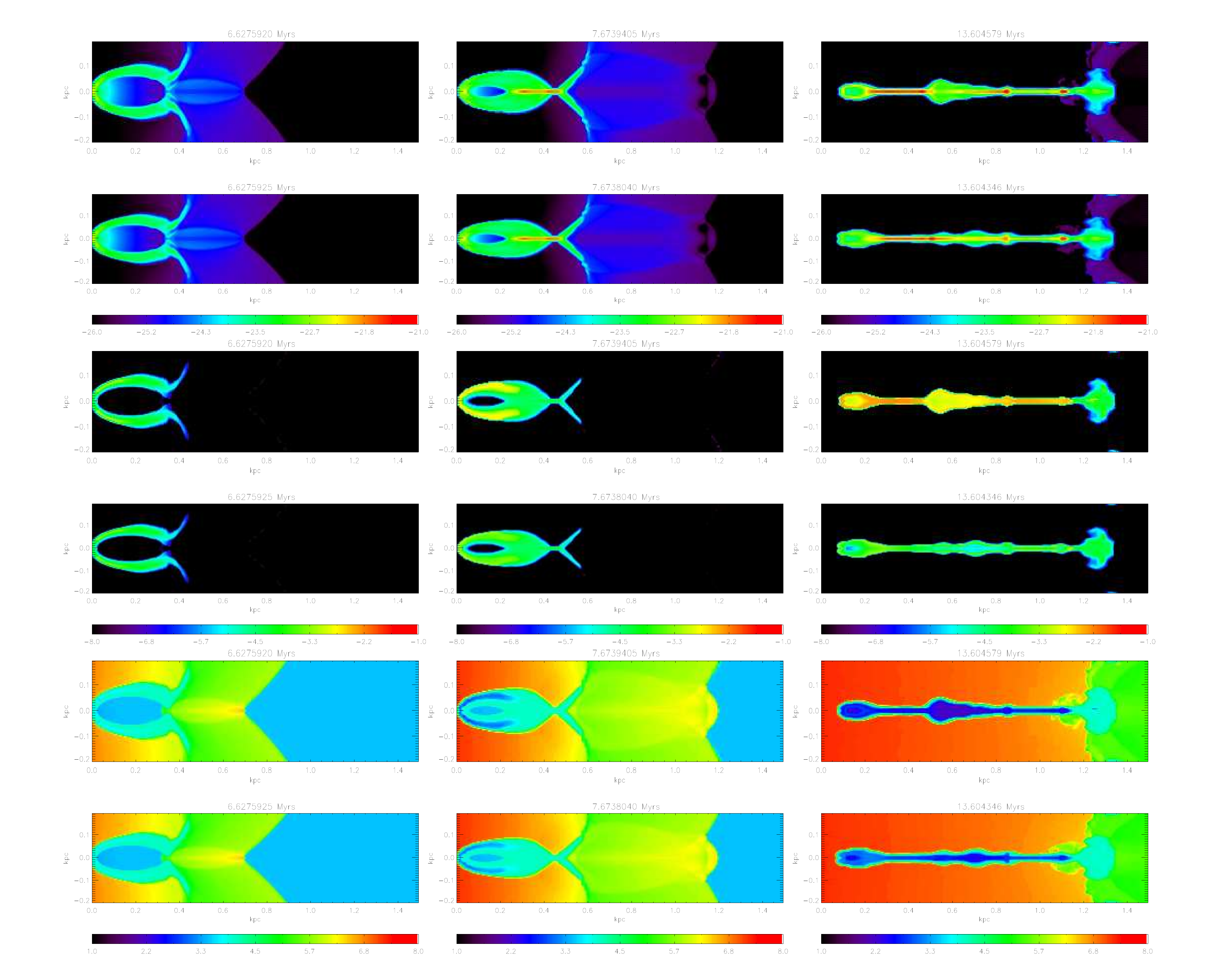}
\caption{Comparison between the fiducial run (HBN) and a run including a dissociating background (HBY) at three important stages of evolution. As in Fig.\  \ref{compcases}, from left to right the columns  correspond to $t=6.6,$ 7.7, and 14.0 Myrs.  From top to bottom the rows represent
log density contours in the fiducial run (Row 1) and the dissociating background run (Row 2), contours of log H$_2$ mass fraction in run HBN (Row 3) and HBY (Row 4), and contours of log temperature from run HBC (Row 5) and HBY (Row 6).  The limits of each panel are the same as Fig~\ref{sim1}.  The addition of a background greatly reduces H$_2$ but has almost no effect on the dynamics of the interaction.}
\label{compuv}
\end{figure*}

A more uncertain aspect of our simulation is our assumption of a negligible dissociating background.  In fact, the presence of at least a low level of dissociating background is necessary in order for the minihalo not to collapse and form stars on its own, cooling by H$_2$ and HD left over from recombination.   To set an upper limit on the impact of such a background we modify the rates in our chemical network to approximate a relatively large dissociating background of J$_{21}= 0.1,$ as discussed in section 2.1.   Furthermore, as these rates are modified for all reactions throughout the simulation,  this background is taken to affect even the densest regions of the cluster.  This is equivalent to assuming that the cloud is optically thin to 11.2 to 13.6 eV photons at all times during the simulation.

Figure~\ref{compuv} shows the comparison between the run including this background (HBY) and the fiducial run HBN. As expected, the abundance of H$_2$ is reduced in the case with the UV background, peaking at about $\approx 10^{-4}$ instead of $\approx 10^{-2}$ in the run without a background. Interestingly, this difference persists even after a few megayears into the simulation, and the abundance of the HPY run remains stable at about $\approx 10^{-4}$. 

However, this value is more than sufficient to cool the gas to the same temperature as in HBN. Even with this lower mass fraction, the cooling time is smaller than the dynamical time, and the evolution of the cloud remains essentially unchanged. The cloud collapses and is stretched into the same configuration as found without a background. Dense clouds are again found between 0.2 and 0.4 kpc, at 0.55 kpc, and 0.9 kpc and the density and temperature of each of these clouds is comparable to those found in the fiducial without a background.  By neglecting any molecular self-shielding, this represents a worst case scenario for H$_2$,  yet shock minihalo-interactions continue to make compact stellar clusters.

\subsection{Resolution}

\begin{figure*}
\centering
\includegraphics[scale=0.45, keepaspectratio=true, trim= 15mm 0mm 0mm 0mm]{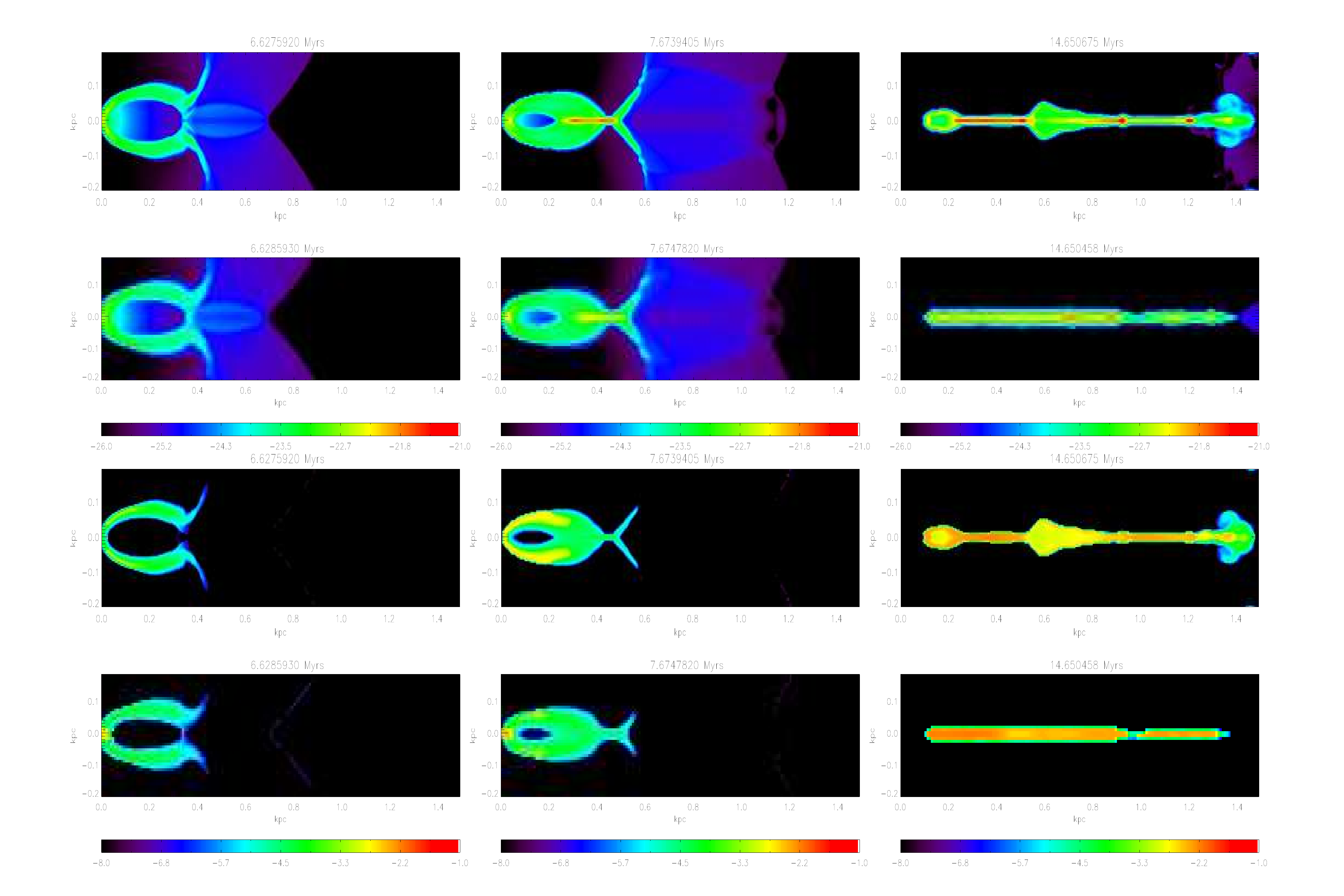}
\caption{Impact of maximum levels of refinement.  Row 1 and Row 3 show the density and H$_2$ contours respectively for the fiducial HBN while Row 2 and Row 4 show contours of density and H$_2$ for LBN. Contour levels are the same as Fig.~\ref{compcases}. Time is given at the top of each panel, and proceeds from $t=6.6$ Myrs (left column) to $t=7.7$ Myrs (center column) to $14.7$ Myrs (right column).}
\label{compres}
\end{figure*}

Finally, we considered the impact of the maximum refinement level on our results. Figure~\ref{compres} compares the fiducial run with 6 levels of refinement and an effective resolution of 4.55 pc to a lower resolution run (LBN) with 5 maximum levels of refinement and an effective resolution of 9.1 pc. Here density is shown in the upper two rows in each pair while the mass fraction of H$_2$ is shown in the bottom two rows.

Only a slight dependence on the formation for H$_2$ is found with resolution. This is most apparent in  the second column, corresponding to $t$ = 7.7 Myrs, which shows that the shocks is slightly broadened in the lower resolution case. Since both chemical reactions and cooling go as $n^2$, this smearing out has the effect of slightly decreasing H$_2$ formation and cooling in the lower resolution run. However, enough H$_2$ is produced in both cases for the cloud to collapse efficiently, and evolve in the same manner up until late times, when the difference in H$_2$ abundance is small.

Furthermore, we also conducted similar resolution studies using Case A recombination, and also modifying chemistry and cooling to account for the presence of a dissociating UV background.  Again comparisons between the high-resolution runs (HAN and HBY) with the low resolution runs (LAN and LBY) uncovered only weak differences with resolution.   Compact stellar clusters were formed in all cases.
 
\subsection{Source of Halo Globular Clusters?}

If indeed shock-minihalo interactions lead to massive clusters of
stars, the longest-lived stars in these objects will remain observable
down to low redshift, providing a direct observational constraint on
our results.  Furthermore, the clusters formed in our simulations are
extremely compact, and thus unlikely to be tidally disrupted as they
eventually become gravitationally bound to the ever larger structures
that form over cosmological time.   As discussed in Scannapieco \etal
(2004), the  most natural candidate for these old and dense stellar
clusters is the population of  metal-poor globular clusters, associated
with the halos of galaxies (\eg Zinn 1985;  Ashman \& Bird 1993;
Larsen \etal 2001; Strader \etal 2005; Brodie \etal 2006).

There are several detailed properties of the clusters in our
simulations that support this association.  Halo globular clusters,
like the more metal-rich (disk) globular clusters, are extremely
compact, with typical half-light radii of $\approx 3$ pc (\eg Jord\'an
\etal 2005).  Unlike higher-metallicity globular clusters, however,
which may have mostly formed at intermediate redshifts (Elmegreen 2010) the
age of all metal-poor globular clusters is between $10-13$ Gyrs,
placing their formation within or before  reionization, during the
epoch in which minihalos had not yet been evaporated by ionization
fronts.

While globular clusters exits at a range of masses, their
mass distribution is well described by a Gaussian in
$\log_{10}(M_{gc})$ with a 0.5 dispersion, and mean at $10^5$ solar
masses, precisely spanning the masses of the clumps formed in our
simulations.  Although the lower mass limit of globular
clusters may be set by destruction through mechanical
evaporation (\eg Spitzer \& Thuan 1972) and shocking that occurs as
the cluster pass through the host galaxies (\eg Ostriker \etal 1972),
the upper mass limit appears to be an intrinsic property of the initial
populations (\eg Fall \& Rees 1985; Peng \& Weisheit 1991;  Elmegreen 2010). 
 Furthermore, this upper limit of $\approx 10^6 M_\odot$ in
stars corresponds roughly
with  the $10^4$ K virial temperature
above which atomic cooling becomes effective in high-redshift
virialized clouds of gas and  dark matter (Fall \& Rees 1985).    No
$\gtrsim 10^6 M_\odot$ stellar clusters can result from shock minihalo
interactions, because no minihalos exist with gas masses $\gtrsim 10^6
M_\odot.$

The second clue as to the origin of globular clusters comes from
observations of stars being tidally stripped from these objects. If,
like galaxies, globular clusters are contained within individual dark
matter halos, the stripping of stars would be highly suppressed, due
to the  increased gravitational potential.  Surprisingly, no such
evidence of dark matter halos is seen (Moore 1995), suggesting that
these clusters formed through a mechanism markedly different than
star formation  within galaxies.

Again, the triggered-star formation seen in our shock-minihalo
simulations provides a natural explanation of this key observed
property.  While gas is initially gathered together by a collapsed dark
matter potential, the lack of effective coolants prevents this 
collapse from continuing to the point that stars are formed at the center
of the potential.  Instead the momentum of the  impinging galaxy wind
is such that it is able to accelerate the gas above the  escape
velocity of the halo --  even as it induces cooling and compresses the
gas to the point that it remains gravitationally bound on its own.  
The trigger for star formation and the
mechanism that removes the dark mater halo are  one and the same,
resulting in a population of stellar clusters that form at the
moment they break free from their dark-matter enclosures.
	
\section{Conclusions}
	
Cosmological minihalos provide the initial building blocks out of which larger dark mater
halos and galaxies form.  During reionization, these minihalos
surrounded  all young galaxies and consisted of
metal-free neutral atomic gas. The lack of molecules in these objects
stalled their evolution at virial
temperature below $\approx 10^4 \rm \ K$ as cooling from atomic hydrogen becomes
inefficient and any molecular coolants are dissociated by UV
Lyman-Werner photons, too hot for star formation in such small
objects. Only interactions that cause non-equilibrium molecule
formation could allow the gas in minihalos to collapse and form stars. 
	
Previous work has used ionization fronts to create these conditions
(Cen 2001), however, 3D hydrodynamic simulations show that for either
a stellar or quasar source, instead of creating clouds of H$_2$, the
cloud is instead completely photo-evaporated (Iliev \etal 2005;
Shapiro \etal 2006). However galactic outflows are another option for triggering
star formation. Shocks provide the conditions for non-equilibrium chemistry without
completely destroying the cloud.
		
To simulate these interactions, we have added a primordial chemistry
network and associated cooling routines into FLASH3.1. 
This network
traces collisional ionization and recombination of hydrogen and helium
as well as the formation of two primary coolants in the absence of
metals: H$_2$ and HD. We have also included the impact of a dissociating
background on these rates and implemented routines that
control the cooling of gas  in the presence atomic and molecular
line cooling.
 
With these code developments in place, we  have been able to 
simulate the interaction between a galactic outflow and a
primordial minihalo that results in structures that are identified as
proto-globular clusters. The shock fills two very important
roles. First, it ionizes the neutral gas found in the minihalo, which recombines and
begins to form  H$_2$ and HD, through nonequillbrium processes.
These coolants allow the cloud to cool to much lower temperatures, triggering star formation. 
Secondly, the shock imparts momentum into the gas and accelerates it 
above the escape velocity. This creates a
cloud of dense, cold molecular gas that is free from dark matter
halos.
	
Together, these many similarities suggest that there is not only room
at low redshifts for the kinds of clusters were see in our
simulations, but that low-redshift observations {\em require} the
formation of stars by a mechanism remarkably similar to the one we are
simulating.   This physical connection requires further investigation,
especially in light of a  final remarkable property of globular
clusters, the chemical homogeneity of their stars.  While there are
large metallicity between different globular clusters the dispersion
of [Fe/H] within  any given cluster  is less than 0.1 dex, or a factor
of $\approx$ 1.25 (Suntzeff 1993).  Usually, this is explained either
by  pre-enrichment,   meaning that stars form out of
gas that had already been homogeneously enriched by a previous
generation of supernovae (\eg Elmegreen \& Efremov 1997; Bromm \&
Clarke 2002),  or ``self-enrichment," meaning that the protocluster
cloud was enriched by one or more SN events occurring within it (e.g.,
Brown \etal 1995; Cen 2001; Nakasato \etal 2000; Beasley et al. 2003;
Li \& Burstein 2003).   Both types of scenarios have strong flaws.  In
the pre-enrichment case,  the key problem is just exactly
what the previous population was, why it played only a secondary role
in the formation history of the cluster, and how it could have
enriched this material on very short timescales. In the
self-enrichment picture, on the other hand, the key problems is that
the kinetic energy corresponding to resulting SN is likely to unbind
the gaseous protocluster (Peng \& Weisheit 1991) long before it
enriches the gas.  Recent two dimensional numerical simulations of primordial
supernova inside minihalos  show that they do in fact unbind these systems over a range
of supernova types and halo sizes (Whalen \etal 2008b).

Shock minihalo interactions have the potential to combine the best
features of both these scenarios, bringing in the metals from an
outside source, such that the gas collapses instead of unbinds, but
nevertheless explaining how star formation could be synchronized in a
large mass of coolant-filled gas.  Yet highly homogenous population of
stars can only be formed if the incoming, enriched gas and the
primordial, minihalo material are sufficiently well mixed before star
formation begins.   Modeling this process will require simulations
that build on the ones here to include metal line cooling above and
below $10^4$ K and detailed models of subgrid-mixing.   Testing this 
assumption and determining what sets of parameters lead to realistic globular 
cluster formation will be the focus of the upcoming papers in this series.

\acknowledgments

We are grateful to Tom Abel, Jean Brodie, Simon Glover, Raul Jimenez, Joaquin Prieto, Jay Strader, Sumner Starrfield, F. X. Timmes, and Steven Zepf for their helpful comments.
ES acknowledges the support from NASA theory grant NNX09AD106.
All simulations were conducted on the ``Saguaro'' cluster operated by the
Fulton School of Engineering at Arizona State University.
The results presented here were produced using the FLASH code, a product of the DOE
ASC/Alliances-funded Center for Astrophysical Thermonuclear Flashes at the
University of Chicago.

\end{document}